\documentclass[a4paper,11pt]{article}
\usepackage{jheppub}

\usepackage{amsmath}
\usepackage{booktabs}
\usepackage{color}

\definecolor{lgris}{rgb}{0.95,0.95,0.95}

\newcommand{\ra}[1]{\renewcommand{\arraystretch}{#1}}
\newcommand{\rb}[1]{\renewcommand{\tabcolsep}{#1}}

\newcommand{\eq}[1]{\begin{equation} #1 \end{equation}}
\newcommand{\eqa}[1]{\begin{eqnarray} #1 \end{eqnarray}}
\newcommand{\nn}{\nonumber}
\newcommand{\nind}{\noindent}
\newcommand{\av}[1]{\langle #1 \rangle}
\newcommand{\C}[1]{{\cal{C}}_{#1}}

\newcommand{\bsg}{\bar B\to X_s\gamma}
\newcommand{\g}{\gamma}
\newcommand{\ord}[1]{{\cal O}(#1)}

\newcommand{\tr}{{\rm Tr}}
\newcommand{\ep}{\epsilon}
\newcommand{\sla}[1]{#1 \!\!\!/}
\newcommand{\sss}{\scriptscriptstyle}
\newcommand{\skipp}[1]{}
\newcommand{\vtwo}[1]{#1}

\title{Four-body contributions to $\bsg$ at NLO} 
\author{Tobias Huber$^a$,} 
\author{Micha\l\ Poradzi\'nski$^{a,b}$,} 
\author{and Javier Virto$^a$} 
\affiliation[a]{Theoretische Physik 1, Naturwissenschaftlich-Technische Fakult\"at, \\
Universit\"at Siegen,  D-57068 Siegen, Germany} 
\affiliation[b]{Faculty of Physics, University of Warsaw, PL-00-681 Warsaw, Poland} 
\emailAdd{huber@physik.uni-siegen.de} 
\emailAdd{michal.poradzinski@fuw.edu.pl} 
\emailAdd{virto@physik.uni-siegen.de} 

\abstract{
Ongoing efforts to reduce the perturbative uncertainty in the $\bsg$ decay rate have resulted in a theory estimate to NNLO in QCD.
However, a few contributions from multi-parton final states which are formally NLO are still
unknown. These are parametrically small and included in the estimated error from higher order corrections,
but must be computed
if one is to claim complete knowledge of the $\bsg$ rate to NLO. A major part of these unknown pieces
are four-body contributions corresponding to the partonic process $b\to s\bar q q\gamma$.
We compute these NLO four-body contributions to $\bsg$, and confirm the corresponding tree-level leading-order results.
While the NLO contributions arise from tree-level and one-loop Feynman diagrams, the four-body phase-space
integrations make the computation non-trivial. The decay rate contains collinear logarithms arising from
the mass regularization of collinear divergences. 
We perform an exhaustive numerical analysis, and find that these contributions are positive and amount to
no more than $\sim 1\%$ of the total rate in the Standard Model, thus confirming previous estimates of the perturbative uncertainty.
}
\preprint{
\begin{minipage}{3cm}
\small
\flushright
QFET-2014-22\\
SI-HEP-2014-29\\
IFT/14/06
\end{minipage}}

\begin{document}

\maketitle

\renewcommand{\thefootnote}{\arabic{footnote}}
\setcounter{footnote}{0}

\section{Introduction}

The inclusive radiative $B$ meson decay $\bsg$ is the paradigm for precision tests of the Standard Model (SM) in quark
flavor physics. Its branching ratio is measured with a precision of $\sim 7\%$
\cite{cleo,belle,babar,HFAG}\footnote{
The semi-inclusive measurement in the first reference in~\cite{belle} has recently been superseded by a new, more precise one --~see Ref.~\cite{1411.3773}. However, this new measurement is not yet taken into account in
the HFAG average of Eq.~(\ref{exp}).
},

\eq{
{\cal B}(\bsg)_{E_\gamma >1.6\text{ GeV}}^\text{exp} = (3.43 \pm 0.22) \cdot 10^{-4}\ .
\label{exp}
}

To match this experimental precision, a
theory calculation to next-to-next-to-leading order (NNLO) accuracy is necessary. This program has been carried out during the last two decades and it is
almost --~but not quite~-- finished. The current theory estimate results in the following value \cite{0609232}:
\eq{
{\cal B}(\bsg)_{E_\gamma >1.6\text{ GeV}}^{\rm \sss SM} = (3.15 \pm 0.23) \cdot 10^{-4}\ ,
}
where the total $\pm 7\%$ uncertainty is due to non-perturbative (5\%), parametric (3\%), higher orders (3\%) and
$m_c$-interpolation ambiguity (3\%) \cite{0609232}.

The calculation can be divided into: 1. Matching conditions
\cite{9308349,9703349,9710336,9710335,9904413,9910220,0007259,0109058,0401041},
2. Anomalous dimensions
\cite{9211304,9304257,9409454,9612313,9711280,9711266,0306079,0411071,0504194,0612329},
and 3. Matrix elements
\cite{9506374,9512252,9602281,9603404,9903305,0105160,0203135,0302051,0505097,0506055,0605009,0607316,0609241,0707.3090,0805.3911,1005.1173,1005.5587,1009.2144,1009.5685,1209.0965}.
Matching conditions and anomalous dimensions are complete up to NNLO since a long time. 
Missing pieces include $m_c$-dependent matrix elements at
NNLO~\cite{0609241,HCMSinprep}, as well as next-to-leading-order (NLO) matrix elements proportional to penguin or CKM-suppressed current-current
operators. The latter are formally NLO but are suppressed by very small Wilson coefficients, and should
indeed be rather small, fitting within the estimated $\sim 3\%$ uncertainty from higher orders~\cite{0104034,0609232}.
However, only a full calculation can provide precise knowledge of their true effect, and we intend to do that here.
This work constitutes part of an ongoing effort to reduce the perturbative uncertainty down to a negligible level.

The $\bsg$ rate is given by the inclusive partonic rate of the $b$ quark, up to non-perturbative effects that,
for a photon energy cut $E_0=1.6$ GeV,
are estimated at the level of $\sim 5\%$ \cite{1003.5012},
\eq{
\Gamma(\bsg)_{E_\gamma>E_0} = \Gamma(b\to X_s^\text{parton} \gamma)_{E_\gamma > E_0} + \ord{1/m_b}
}
where $b\to X_s^\text{parton} \gamma$ denotes the partonic decay of the $b$ quark into any number of particles with an overall strangeness $S=-1$, plus a hard photon with $E_\gamma>E_0$, and \emph{excluding} charm:
\eq{
\Gamma(b\to X_s^\text{parton}\gamma) = \Gamma(b\to s\gamma) + \Gamma(b\to s g\gamma)
+ \Gamma(b\to sq\bar q\gamma) + \cdots \, ,
}
with $q=u,d,s$. Each individual rate is an interference of different amplitudes, computed as
matrix elements of local operators in the effective Lagrangian:
\eq{
{\cal L}_\text{eff} = {\cal L}_{\rm \sss QED+QCD} + \frac{4 G_F}{\sqrt2} V_{ts}^* V_{tb}
\left[  \sum_{i=1}^2 (\C{i}^u P_i^u + \C{i}^c P_i^c) + \sum_{i=3}^8 \C{i} P_i  \right] + h.c.\ ,
}
where the operators $P_i$ are defined as \cite{9612313}:
\eq{
\begin{array}{rclrcl}
P_1^q &=& (\bar s_L \gamma_\mu T^a q_L)(\bar q_L \gamma^\mu T^a b_L)\ ,
& P_2^q &=& (\bar s_L \gamma_\mu q_L)(\bar q_L \gamma^\mu b_L)\ ,\\[1mm]
P_3 &=& (\bar s_L \gamma_\mu b_L) \sum_q(\bar q \gamma^\mu q)\ ,
& P_4 &=& (\bar s_L \gamma_\mu T^a b_L) \sum_q(\bar q \gamma^\mu T^a q)\ ,\\[1mm]
P_5 &=& (\bar s_L \gamma_\mu\gamma_\nu\gamma_\rho b_L) \sum_q(\bar q \gamma^\mu\gamma^\nu\gamma^\rho q)\ ,
& P_6 &=& (\bar s_L \gamma_\mu\gamma_\nu\gamma_\rho T^a b_L) \sum_q(\bar q \gamma^\mu\gamma^\nu\gamma^\rho T^a q)\ ,\\[1mm]
P_7 &=& (e/16\pi^2) m_b (\bar s_L\sigma^{\mu\nu} b_R)F_{\mu\nu}\ ,
& P_8&=& (g_s/16\pi^2) m_b (\bar s_L\sigma^{\mu\nu} T^a b_R)G^a_{\mu\nu}\ .
\end{array}
\label{eq:ops}
}
With this notation, $\C{1,2}^{q}$ contain CKM phases:
$\C{1,2}^{q} = -\lambda_q C_{1,2}$, with $\lambda_q\equiv V_{qs}^* V_{qb}/V_{ts}^* V_{tb}$ and $C_{1,2}$ defined in the usual way, e.g.\ Ref.~\cite{9612313}. We will also use the notation
$\C{1u}\equiv \C{1}^u$ etc. The other Wilson coefficients are simply $\C{3,..,8}=C_{3,..,8}$
as in Ref.~\cite{9612313}.

For more than two final state particles, the rate involves integration over phase space; the photon spectrum opens up, and the rate depends on the photon-energy
cut. The perturbative contribution can be written, in the notation of Ref.~\cite{0506055}, as:
\eq{
\Gamma(b\to X_s^\text{parton} \gamma)_{E_\gamma > E_0} = \Gamma_0\, \sum_{i,j} \C{i}^\text{eff}(\mu_b)^*\,\C{j}^\text{eff}(\mu_b)\, \widetilde G_{ij}(\mu_b,\delta)\ ,
\label{bsgrate}
}
summed over $i,j = 1u,2u,3,..,6,1c,2c,7,8$, and with the normalization
\eq{\Gamma_0 = \frac{G_F^2 m_b^5 \alpha_e |V_{ts}^*V_{tb}|^2}{32\pi^4}\ .}
The ``effective'' Wilson coefficients are $\C{1q,2q,3,..,6}^\text{eff} = \C{1q,2q,3,..,6}$,
$\C7^\text{eff} = \C7- \frac13 \C3 - \frac{4}{{\vtwo{9}}} \C4 - \frac{20}3 \C5 - \frac{80}9 \C6$ and
$\C8^\text{eff} = \C8+\C3 - \frac16 \C4 + 20 \C5 - \frac{10}3 \C6$.
Throughout the paper we use the NDR-$\overline{\rm MS}$ scheme with fully anti-commuting $\gamma_5$. 

The objects $\widetilde G_{ij}$ arise from the interference of operators $P_i$ and $P_j$ in the squared matrix elements, integrated
over phase space. They depend on the photon energy cut through the variable
$\delta\equiv 1-2E_0/m_b$.
In the notation of Ref.~\cite{0609241}, where normalization to the inclusive semileptonic $b\to u$ rate is used,
$K_{ij} = \widetilde G_{ij}/G_u$, with
$G_u=1+C_F(\frac{25}2-12\zeta_2)\frac{\alpha_s}{4\pi} + \ord{\alpha_s^2}$~\cite{9903226}.

In this paper we focus on the four-body contributions to $\Gamma(b\to X_s^\text{parton} \gamma)_{E_\gamma > E_0}$, corresponding
to $\Gamma(b\to s\bar q q \gamma)$:
\eq{
\Gamma(b\to s\bar q q \gamma)_{E_\gamma > E_0} = \Gamma_0\, \sum_{i,j} \C{i}^\text{eff}(\mu_b)^*\,\C{j}^\text{eff}(\mu_b)\, G_{ij}(\mu_b,\delta)\ ,
\label{bsqqgrate}
}
where we define $G_{ij}$ as the $b\to s\bar q q \gamma$ contribution to $\widetilde G_{ij}$.
In addition, we expand $G_{ij}$ as:
\eq{
G_{ij}(\mu,\delta) = G_{ij}^{(0)}(\delta) + \frac{\alpha_s(\mu)}{4\pi} G_{ij}^{(1)}(\mu,\delta)
+\ord{\alpha_s^2}\ .
}

\begin{figure}
\centering
\includegraphics[width=\textwidth]{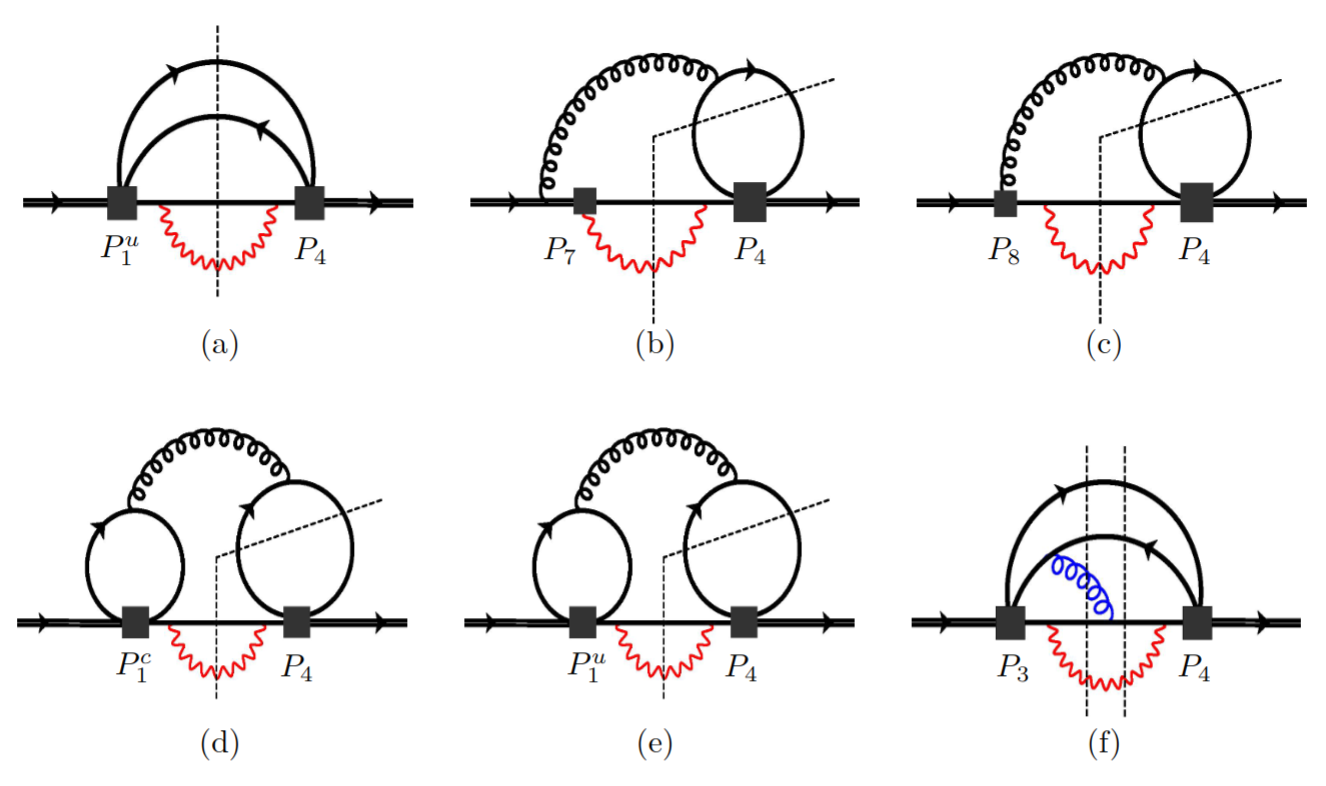}
\vspace{-7mm}
\caption{Sample cut-diagrams contributing to LO and NLO four-body matrix elements. In our calculation
we include all contributions but those from panel (f). See the text for details.}
\label{fig:examples}
\end{figure}

There is a hierarchy in the size of the Wilson coefficients of the operators in Eq.~(\ref{eq:ops}), which
can be divided into two classes:
\eq{
A=\{P_1^c,P_2^c,P_7,P_8\}\ ; \quad B=\{P_1^u,P_2^u,P_3,P_4,P_5,P_6\}\ .
}
Operators in class $A$ have large Wilson coefficients, whereas class-$B$ operators have either very small
Wilson coefficients or are CKM-suppressed. Among the four-body leading and next-to leading contributions we distinguish four groups:

\begin{itemize}

\item Tree-level $(B,B)$ interference (Fig.~\ref{fig:examples}.a). These are the leading-order (LO) contributions and provide the complete matrix $G^{(0)}(\delta)$. This matrix has been computed in Ref.~\cite{1209.0965}.

\item Tree-level $(A,B)$ interference (Figs.~\ref{fig:examples}.b and \ref{fig:examples}.c). These are
NLO and provide the matrix elements $G_{7j}^{(1)}$ and $G_{8j}^{(1)}$, with $j=1u,2u,3,..,6$.

\item One-loop $(A,B)$ interference (Fig.~\ref{fig:examples}.d). These are NLO and provide the matrix elements $G_{ij}^{(1)}$, with $i=1c,2c$ and  $j=1u,2u,3,..,6$.

\item One-loop $(B,B)$ interference (Figs.~\ref{fig:examples}.e and \ref{fig:examples}.f). The ones in
panel (e) can be obtained from the ones in panel (d) as discussed in Section~\ref{sec:OpId}, and
provide the matrix elements $G_{ij}^{(1)}$, with $i,j=1u,2u,3,..,6$. 
The ones in panel (f) include five-particle cuts since the one-loop four-body diagrams must be complemented with the five-body gluon-bremsstrahlung correction $b\to s\bar q q \gamma + g$. We therefore leave the contributions from panel (f) aside from the present four-body calculation.

\end{itemize}

We note that NLO $(A,B)$ interference terms are presumably as large as the $(B,B)$ interference at the LO since $\C{1u,2u,3,..,6} \sim \alpha_s/\pi\,\C{1c,2c,7,8}$. For the same
reason, the partly neglected $(B,B)$ interference terms at the NLO are expected to be much smaller than the $(A,B)$ interferences that we calculate in a complete manner. As a final comment, we note that four-body NNLO contributions of the type $b\to sgg\gamma$ are part of the calculation in Ref.~\cite{HCMSinprep}, and do not interfere with our results.

The structure of the paper is the following. In Section~\ref{sec:2} we discuss the details of our calculation and the structure
of the different contributions, including the set of diagrams needed, the UV renormalization, and the treatment of
collinear divergences. In Section~\ref{sec:results} we collect the final results. In Section~\ref{sec:numerics}
we perform a numerical study of all the evaluated interferences. Section~\ref{sec:conclusions} contains our conclusions. In
Appendix~\ref{sec:interresults} we collect some intermediate results of the calculation, where analytic cancellation of UV and collinear
divergences can be explicitly checked.

\section{Details of the calculation}
\label{sec:2}

The NLO calculation is performed in 4 steps:

\begin{enumerate}

\item Evaluation of the cut-diagrams shown in Figs.~\ref{diagsi}, \ref{diagsii}, \ref{diagsiii}. We use the
Cutkosky rules for cut (on-shell) propagators, accounting for spin and color sums for all external particles.
The result of each diagram is a contribution to the differential decay rate
${\cal K}(s_{ij})$, a scalar function  of the momentum invariants $s_{ij}$, $i,j=1,..,4$, with $i\ne j$ (see
Sections~\ref{sec:PSint} and~\ref{sec:epsterms}).

\item Integration over the four-particle phase-space. This step is described in Section~\ref{sec:PSint}.

\item Renormalization: This requires the evaluation of the tree-level diagrams with counterterms shown in
Fig.~\ref{fig:counterterms}, and the corresponding phase-space integration. As a bonus, this step allows one to
check the LO result for $G_{ij}^{(0)}$ of Ref.~\cite{1209.0965}. This step is described in
Section~\ref{sec:renormalization}.

\item Collinear logarithms: Having regularized collinear divergences in dimensional regularization, we use
the method of splitting functions \cite{1209.0965,9605323,0011222,0107138,0301047} to transform $1/\epsilon_\text{coll}$
poles into collinear logarithms of the form $\log(m_q/m_b)$. This requires the computation of the corresponding
$b\to s\bar q q$ corrections with subsequent photon emission $q'\to q'\gamma$ (with $q'=q,s$), by evaluation of the
diagrams shown in Fig.~\ref{fig:collinear}, the convolution with the splitting function,
and the three-particle phase-space integration. This step is described in Section~\ref{sec:collinear}.

\end{enumerate}

We note that every diagram has to be computed inserting all the operators $P_{1u,2u,3,..,6}$ to the right
of the cut, and $P_{1u,2u,1c,2c,3,..,8}$ to the left (see e.g.\ Fig.~\ref{fig:examples}),
leading to all the different interference terms in the matrix
$G_{ij}^{(1)}$. In Section~\ref{sec:OpId} we argue that most of the elements of this matrix can be obtained
from a reduced set by the use of different operator relations and Fierz identities. In addition, this spells out
transparently the dependence of the full result on the Wilson coefficients before any calculation is performed.
We will see that --~with one exception discussed at the end of Section~\ref{sec:OpId-Left}~--
only diagrams with $P_{7,8,1c}$ to the left of the cut and
$P_4$ to the right must be considered. This simplifies the calculation considerably.

Finally, for each diagram in  Figs.~\ref{diagsi}$-$\ref{fig:collinear},
there is the corresponding mirror image. These ``mirror'' contributions are just obtained
by complex conjugation, and ensure that the rate is real, i.e.\ $G_{ij}^{(1)}$ is hermitian. We disregard these
mirror contributions in the calculation, but at the end we substitute
$G_{ij}^{(1)}\to G_{ij}^{(1)} + G_{ji}^{(1)\,*}$.

\subsection{Operator identities}
\label{sec:OpId}

\subsubsection{Color}

Diagrams with insertion of the color octet operators $P_{4,6}$ are related 
to the ones with insertion of color singlet operators $P_{3,5}$ by a simple color factor, 
which can be obtained just from the color structure of the gluon penguin:
\eq{
\raisebox{-8mm}{
\includegraphics[width=4.6cm]{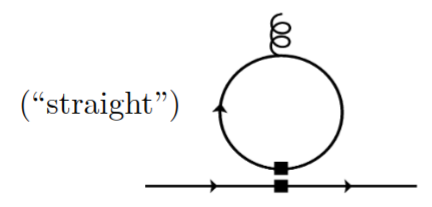}}
\quad \longrightarrow\quad \left\{
\begin{array}{ll}
tr(T^a)=0 & \text{color singlet}\\[2mm]
tr(T^b T^a) T^b = \frac12 T^a \quad & \text{color octet}
\end{array} \right.
}
\eq{
\raisebox{-8mm}{
\includegraphics[width=4.6cm]{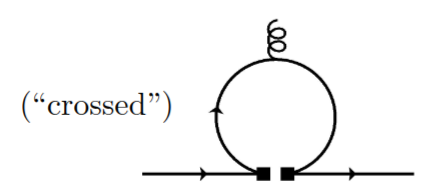}}
\quad \longrightarrow\quad  \left\{
\begin{array}{ll}
T^a & \text{color singlet}\\[2mm]
T^b T^a T^b = -\frac1{2N_c} T^a \quad & \text{color octet}
\end{array} \right.
}\\
This leads to the rule that $P_{3,5}$ can always be replaced by:
\eqa{
P_{3,5} &\to& 0  \hspace{1.9cm} \text{(straight insertion)}\ ,\label{colorS}\\
P_{3,5} &\to& -6\,P_{4,6}  \qquad \text{(crossed insertion)}\ . \label{colorx}
}
For the same reason, one can always put $P_2^q\to -6 P_1^q$, 
meaning that $\C{1,2}^q$ always come in the combination $(\C1^q-6\,\C2^q)$.

\subsubsection{Insertions to the right of the cut}
\label{sec:OpId-Right}

We restrict ourselves here to the insertion of operators to the \emph{right} of the cut.
Using the 4D identity 
$\g^\mu \g^\nu \g^{\lambda} = g^{\mu\nu}\g^{\lambda} + g^{\nu\lambda}\g^\mu - 
g^{\mu\lambda}\g^\nu + i\epsilon^{\mu\nu\lambda\alpha}\g^\alpha\g_5$ 
we find:
\eq{P_6 = 10 P_4 - 6 \widetilde P_4 + \ord\ep\label{eq:P6P4P4tilde}\ ,}
where $\widetilde P_4 = \sum_q(\bar s_L\g^\mu T^a b_L)(\bar q \gamma_\mu \gamma_5 T^a q)$. 
We now consider the following ``crossed'' insertion of $\widetilde P_4$ 
into a massless fermion loop with an arbitrary number of \emph{vector} currents:
\eq{
\begin{minipage}{3cm}
\centering
\includegraphics[height=2cm,width=3cm]{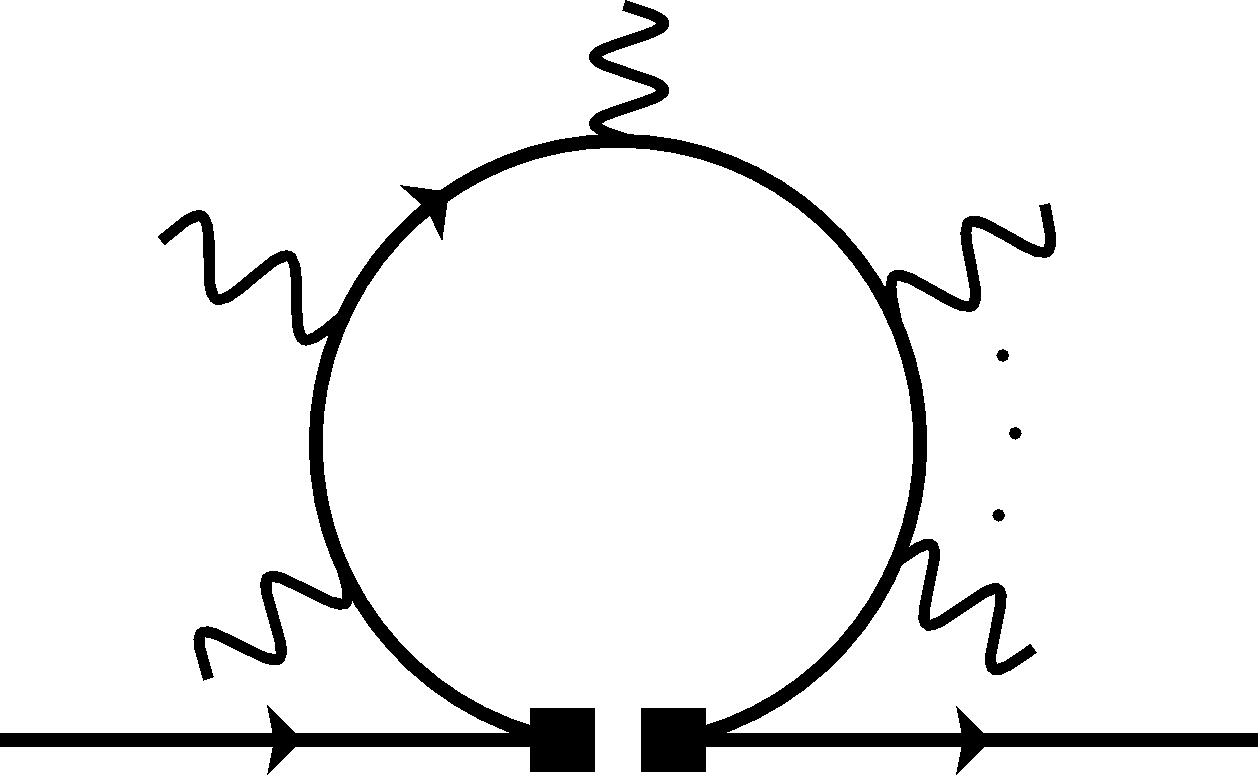}\\
$\widetilde P_4$
\end{minipage}
\quad = \cdots \g_\mu P_L \cdot \underbrace{\g^{\mu_1}\g^{\mu_2}\cdots\g^{\mu_N}\cdot \g^\mu}_{\text{even \# of $\g$'s}} \gamma_5 \cdots = \quad
\begin{minipage}{3cm}
\centering
\includegraphics[height=2cm,width=3cm]{figs/diag1.jpg}\\
$- P_4$
\end{minipage}\label{eq:P4P4tilde}
}
There is always an even number of gamma matrices to the left of $\g_5$,
which can be moved besides the projector $P_L$. 
After that, the relationship $P_L \g_5=-P_L$ provides the given negative sign. 

The ``straight'' insertion of $\widetilde P_4$ does not vanish in general but does not contribute in our case: In the situation where \emph{one} vector current is attached to the loop (Fig.~\ref{diagsi}), the result is proportional to $\tr[\g^\mu\g^\nu\g^\rho\g^\lambda\g_5]\sim \ep^{\mu\nu\rho\lambda}$, but there are not enough independent momenta to be contracted with the antisymmetric tensor, so this contribution vanishes. This is true also in the case where the photon couples twice to the quark loop (Fig.~\ref{diagsiii}). In the case of \emph{two} current insertions (Fig.~\ref{diagsii}) the result is non-zero, but summing over $u,d,s$ quarks in the loop the result is proportional to $Q_u+Q_d+Q_s=0$.

Summing up, in the diagrams with a $P_6$ insertion one can always substitute:
\eqa{
P_6 &\to & 10 P_4 + \ord\ep \qquad \text{(straight insertion)}\ ,\nn\\
P_6 &\to & 16 P_4 + \ord\ep \qquad \text{(crossed insertion)}\ .\label{P6P4ruleX}
}
The replacement rules~(\ref{P6P4ruleX}) combined with Eqs.~(\ref{colorS}) and (\ref{colorx}) imply 
that the full contribution from $P_{3,4,5,6}$ can be obtained from the terms proportional to $\C4^*$:
\eqa{
\hspace{-10mm}\text{Result(}P_{3,4,5,6}\text{,straight)}&=& ( \C4^*  + 10\,\C6^* ) \times
\text{Result($P_4$,straight)},\label{eq:resultStraight}\\[2mm]
\hspace{-10mm}\text{Result(}P_{3,4,5,6}\text{,crossed)}&=& (-6\,\C3^* + \C4^* - 96\,\C5^* + 16\,\C6^* ) \times
\text{Result($P_4$,crossed)}. \label{eq:resultCrossed}
}
Since this is based on a 4D identity, it is in principle only true up to evanescent terms. Below we show that up to the needed order in $\ep$ these terms do not contribute. We have also checked this by direct computation.

The (crossed) insertion of the operators $P_{1,2}^u$ can also be obtained from $P_4$ by an argument almost identical to that of Eq.~(\ref{eq:P4P4tilde}). In this case one must pay attention to the case where the photon couples to the loop, where the $P_4$ and $P_{1,2}^u$ contributions are proportional to different charge factors.

\subsubsection{Insertions to the left of the cut}
\label{sec:OpId-Left}

We have shown that we only need to compute diagrams with an insertion of $P_4$ to the right of the cut. To the 
\emph{left} of the cut we must insert $P_{7,8}$ as well as $P_{3,4,5,6}$ and $P_{1,2}^{u,c}$. As before, $P_{2,3,5}$ contributions are related to $P_{1,4,6}$ by a simple color factor. In addition, the contribution from $P_1^u$ is obtained from the $P_1^c$ insertion with the replacement $m_c\to 0$. We now show that insertions of $P_{4,6}$ are also known from the insertions of $P_7$, $P_8$ and $P_1^c$.

First we consider the case of the photon penguin, where the gluon does not couple to the fermion loop to the left of the cut. By direct inspection be find that:
\eq{
\left.
\raisebox{-10mm}{
\includegraphics[width=3.5cm]{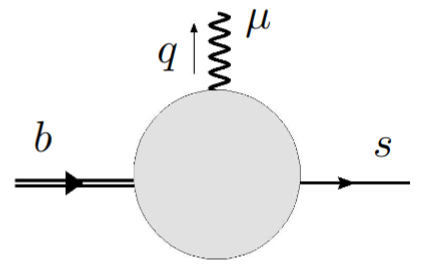}}
\hspace{4mm}
\right|_{q^2=0}
= -\frac{i e}{(4\pi)^2} (\C7^\text{eff}+\ord{\epsilon})  m_b [\sla q,\gamma^\mu] P_R + X\sla{q}q^\mu P_L\ ,
}\\[2mm]
where $\C7^\text{eff}=\C7 -\C3/3-4\,\C4/{\vtwo{9}}-20\,\C5/3-80\,\C6/9$ is the usual ``effective" Wilson coefficient \cite{9711280}, which includes the contributions from $b$-quark loops.
The $\ord{\epsilon}$ corrections are irrelevant to our calculation as the contributions from $P_7$ are finite. 
The term $X\sla{q}q^\mu$ denotes the contribution from four-quark operators proportional to the structure $[\sla{q}q^\mu-q^2 \gamma^\mu]$. This last term does not contribute in our case. To see this, consider the insertion of $P_{1,2}$ into the full diagram:
\eq{
\raisebox{-10mm}{
\includegraphics[width=3.8cm]{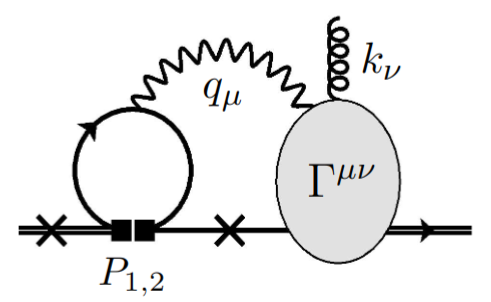}}
\hspace{0.5cm} \sim   \cdots \Gamma^{\mu\nu}\cdots (\sla{q} q_\mu - q^2\g_\mu)P_L\cdots =0\ .
}\\
Here the gluon is attached to either `$\times$'. Since we cut the photon propagator, the photon is on-shell but there is no $q^2$ denominator, and therefore the $q^2$ term cancels. The term $\sla{q} q_\mu$ also cancels due to the Ward identity $q_\mu \Gamma^{\mu\nu}=0$. Note that non-zero contributions to the Ward identity vanish since they either involve a massless fermion tadpole, or if the gluon couples to the loop then it does not depend on the incoming/outgoing quark momenta. We have also checked this result by explicit computation, and indeed the different sets of diagrams satisfying the Ward identity vanish identically.

To summarize: all contributions from photon penguins to the left of the cut are obtained from the diagrams with
insertion of $P_7$ by the replacement $\C7\to \C7^\text{eff}$.

Next, we consider the case of the gluon penguin, where the photon does not couple to the fermion loop to the left of the cut. We find:
\eq{
\raisebox{-11mm}{
\includegraphics[width=3.7cm]{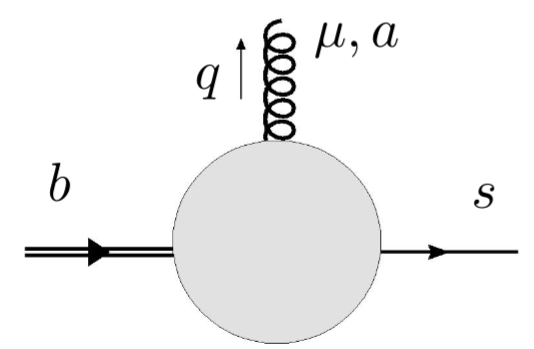}\\[-5mm]}
= -\frac{i g_s}{(4\pi)^2} \Big[ (\C8^\text{eff} +\ord{\epsilon}) m_b T^a [\sla q,\gamma^\mu] P_R  + X T^a (\sla{q}q^\mu-q^2\gamma^\mu) P_L\Big]\ ,
}\\
where $\C8^\text{eff}=\C8 +\C3-\C4/6+20\,\C5-10\,\C6/3$, as usual (e.g.\ Ref.~\cite{0203135}). As before, $\ord{\epsilon}$ corrections to the chromomagnetic
contribution are irrelevant for our calculation because contributions from $P_8$ are UV-finite (collinear divergences are inconsequential here). In the last term, the quantity $X$ is given by:
\eqa{
X&=& -\frac16 \bigg[ (\C1^c-6\,\C2^c)\,g(m_c) + (\C1^u-6\,\C2^u)\,g(0) + (\C4-6\,\C3) [g(0) + g(m_b)]\nn\\[3mm]
&&+  (4\,\C6-24\,\C5) (4-\epsilon-\epsilon^2) \,[g(0) + g(m_b)]\nn\\[3mm]
&&- \frac{6\,\C4 +(60-36\epsilon)\C6}{1-\epsilon}\,[n_\ell\,g(0) + g(m_c) + g(m_b)] \bigg]\ ,
}
where $n_\ell=3$ is the number of light flavors and 
\eq{
g(m) = \frac23 (1-\epsilon) \mu^{2\epsilon} e^{\epsilon\gamma_E} m^{-2\epsilon}\, \Gamma(\epsilon)\,{}_2F_1(\epsilon,2;5/2;q^2/4m^2)
\label{eq:g(m)}}
is the loop integral to all orders in $\epsilon$. Therefore, all contributions from gluon penguins to the left of the cut are known from the contribution of $P_8$
and $P_1^c$.

Now we consider the case in which both the photon and the gluon couple to the loop:
\eqa{
\includegraphics[width=13cm]{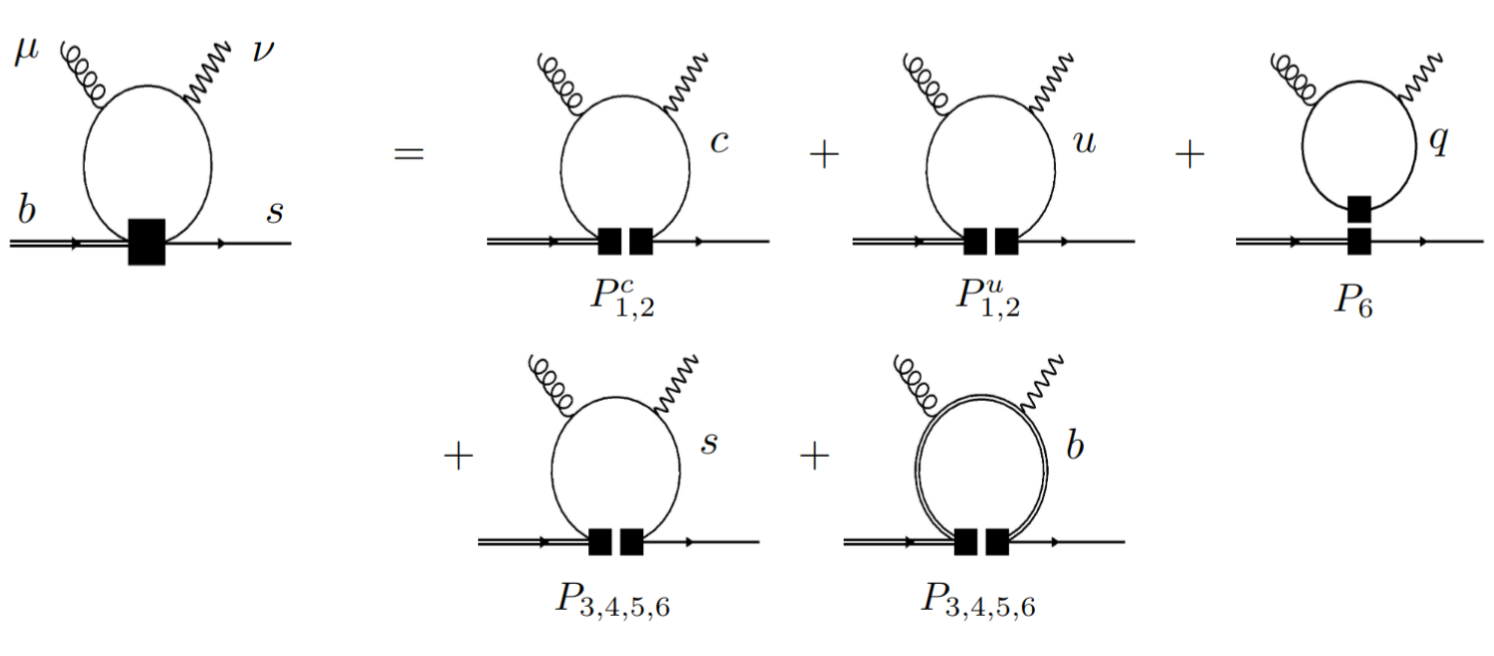}
\nonumber\\[-20mm]
\label{GluonPhoton}
\\[5mm]
\nonumber
}
When inserted into the full diagrams,  these contributions are both UV-finite and collinear safe\footnote
{
We consider always the sum of the two diagrams obtained by swapping the gluon and photon insertions.
}, and we can use 4D identities. The first term in the right-hand side of
Eq.~(\ref{GluonPhoton}) can be written as $Q_u (\C1^c - 6\C2^c) h^{\mu\nu}(m_c)$, which defines the function $ h^{\mu\nu}(m)$.
The second term can be obtained from the first term by the replacement $m_c\to0$: $Q_u (\C1^u - 6\C2^u) h^{\mu\nu}(0)$.
The third term ($q=u,d,s,c,b$) contains only the insertion of $P_6$, because $P_{3,5}$ insertions are
zero due to color, while the insertion of $P_4$ vanishes due to Furry's theorem. For $P_6$ we can make use of the Fierz identity\footnote{Here we use the notation
$P_3^u=(\bar s_L\gamma_\mu b_L)(\bar u\gamma^\mu u)$, etc.}:
\eq{P_1^q = -\frac{4}{27}P_3^q+\frac{1}{9}P_4^q+\frac{1}{27}P_5^q-\frac{1}{36}P_6^q + \ord{\epsilon}\ ,}
which implies that we can trade the straight insertion of $P_6^q$ with the crossed insertion of $-36 P_1^q$. Note also that the contributions from $q=u,d,s$ add up to zero
in the massless limit due to electric charge: $Q_u+Q_d+Q_s=0$. This means that the third term in Eq.~(\ref{GluonPhoton}) is
given by $-36\,\C6 (Q_u h^{\mu\nu}(m_c)+ Q_d h^{\mu\nu}(m_b))$.

The fourth term can be obtained from the first one using the identities in Eqs.~(\ref{eq:P6P4P4tilde}) and (\ref{eq:P4P4tilde}), leading to:
$Q_d(-6\,\C3+\C4-96\,\C5+16\,\C6)h^{\mu\nu}(0)$. The fifth term cannot be completely determined from the insertion of $P_1$ due to the chirality structure.
Using the Fierz identity (\ref{eq:P6P4P4tilde}) we can trade $P_6^b\to 4 P_4^b+12P_1^b$. The second piece, together with the corresponding contribution from $P_5$, results in
$Q_d (-72\,\C5 + 12\,\C6) h^{\mu\nu}(m_b)$. The rest will provide a term $Q_d(-6\,\C3+\C4-24\,\C5+4\,\C6) \widetilde h^{\mu\nu}(m_b)$, where
$\widetilde h^{\mu\nu} \ne h^{\mu\nu}$. Altogether, the right-hand side of Eq.~(\ref{GluonPhoton}) can be written as:
\eqa{
&&Q_u (\C1^c - 6\C2^c) h^{\mu\nu}(m_c) + Q_u (\C1^u - 6\C2^u) h^{\mu\nu}(0) - 36\,\C6 Q_u h^{\mu\nu}(m_c)\nn\\
&&-Q_d (72\,\C5 +24\,\C6) h^{\mu\nu}(m_b) +Q_d(-6\,\C3+\C4-96\,\C5+16\,\C6)h^{\mu\nu}(0)\nn\\
&&+Q_d(-6\,\C3+\C4-24\,\C5+4\,\C6) \widetilde h^{\mu\nu}(m_b)\ .
\label{eq:3charm}}

Therefore only diagrams with insertions
of the operators $P_{7,8}$ and $P_1^c$ to the left of the cut need to be calculated, plus the extra contribution from $\widetilde h^{\mu\nu}(m_b)$. All these relationships
shape the structure of the full results displayed below in the following sections.

\subsection{Set of Diagrams}
\label{sec:Diags}

There are three types of diagrams:\\

\begin{figure}
\centering
\includegraphics[height=3.2cm,width=4.5cm]{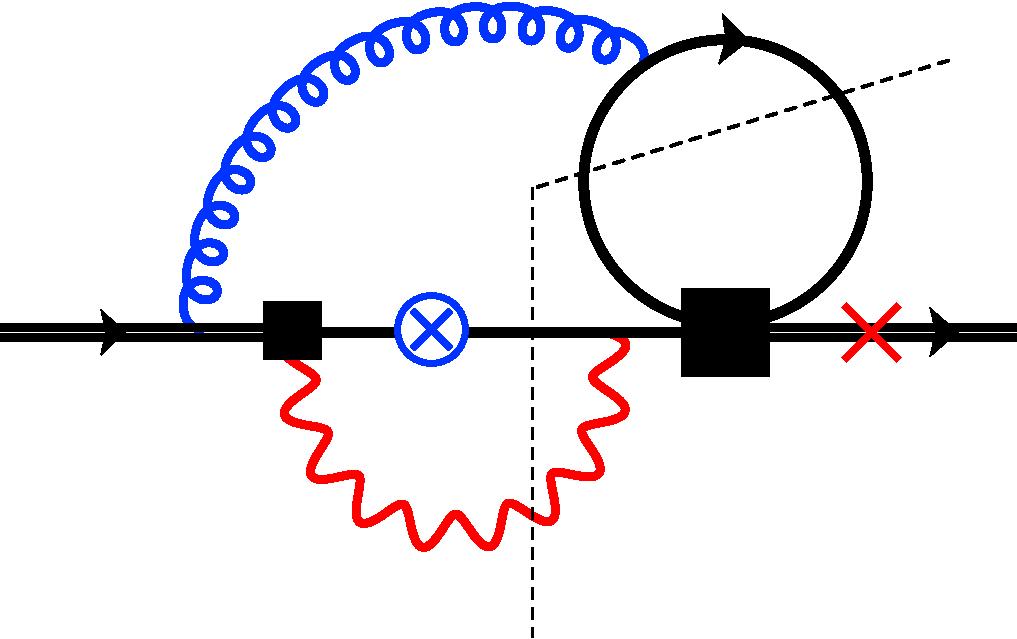}\hspace{5mm}
\includegraphics[height=3.2cm,width=4.5cm]{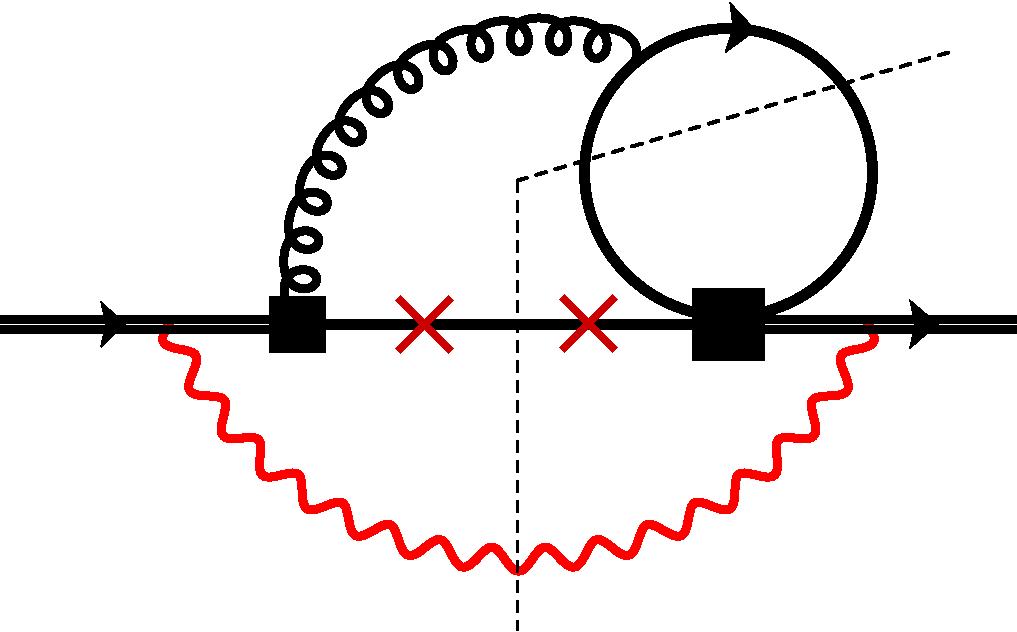}\hspace{5mm}
\raisebox{1.5mm}{\includegraphics[height=3.2cm,width=4.2cm]{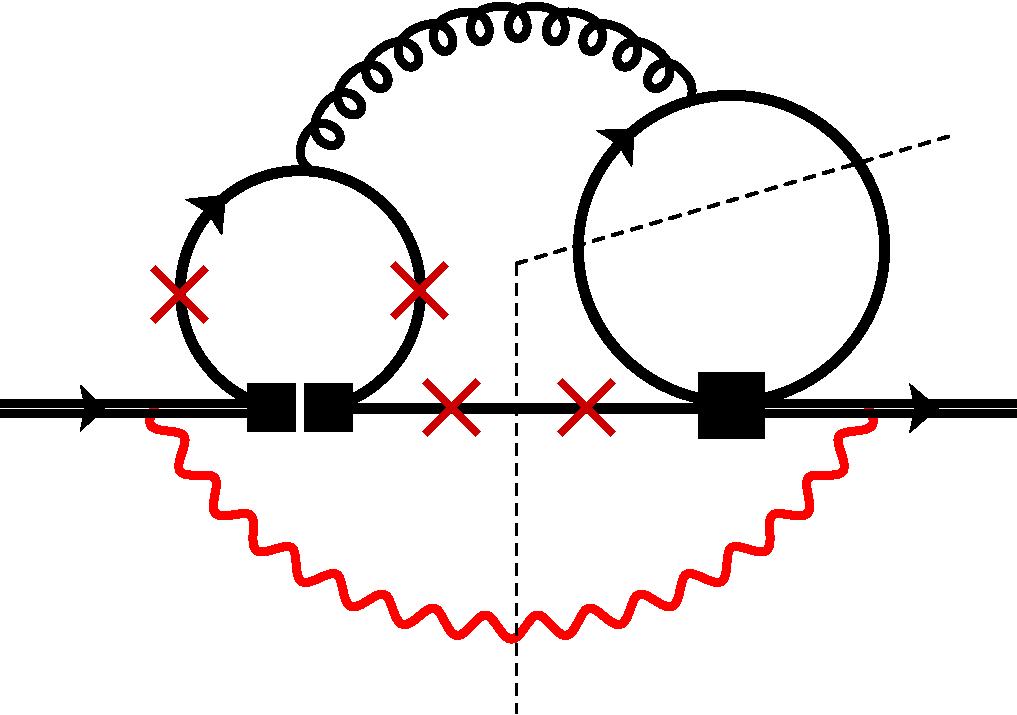}}
\caption{Diagrams of type (i).  Crosses denote alternative insertions of the photon vertex (always one vertex at each side of the cut). Circle-cross denotes the alternative insertion of the gluon vertex.}
\label{diagsi}
\end{figure}

\nind {\bf Type (i):} The photon does not couple to the cut fermion loop (Fig.~\ref{diagsi}): 
In this case crossed and straight diagrams contribute. 
In addition, straight diagrams contain a factor $n_\ell$. 
All in all, the contribution from these diagrams is:
\eq{
{\cal D}_{(i)}^k = Q_d \,\left[ n_\ell\,(\C4^* + 10\,\C6^*)\av{P_4}^{s,k}_{(i)}
+(\C1^{u*}\!-6\,\C2^{u*}\!-6\,\C3^*+\C4^*-96\,\C5^*+16\,\C6^*)\av{P_4}^{\times,k}_{(i)} \right]\ ,
\label{eq:(i)}}
where $\av{P_4}^{s,k}_{(J)}$ and $\av{P_4}^{\times,k}_{(J)}$  denote the contributions to $(P_k,P_4)$ interference terms
from straight and crossed insertions of the operator $P_4$ to the right of the cut, to diagrams of type~$(J)$, respectively.\\ 

\begin{figure}
\centering
\includegraphics[height=3cm,width=4.5cm]{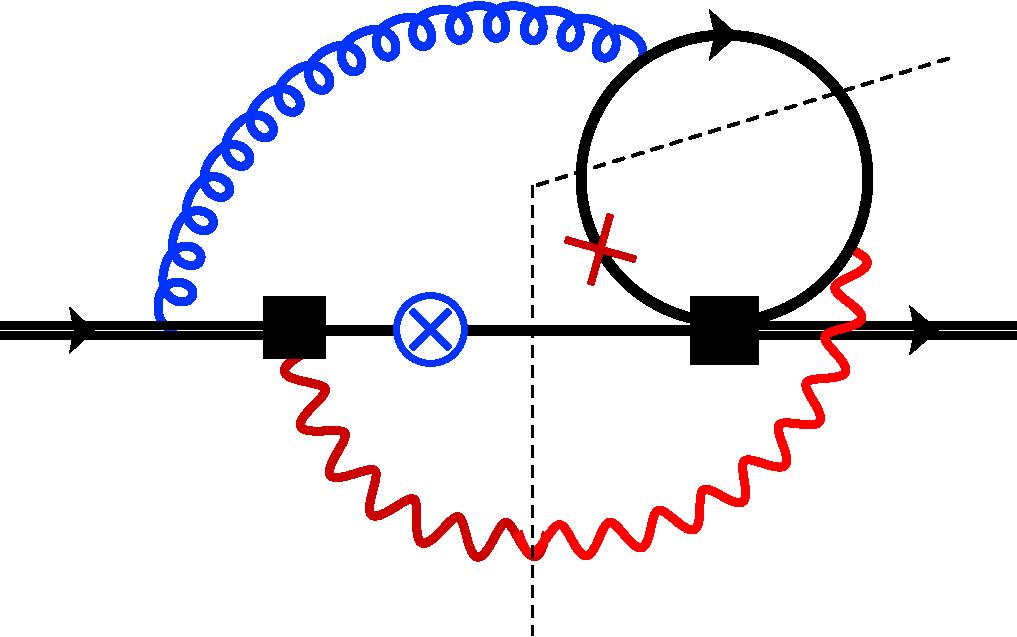}\hspace{5mm}
\includegraphics[height=3cm,width=4.5cm]{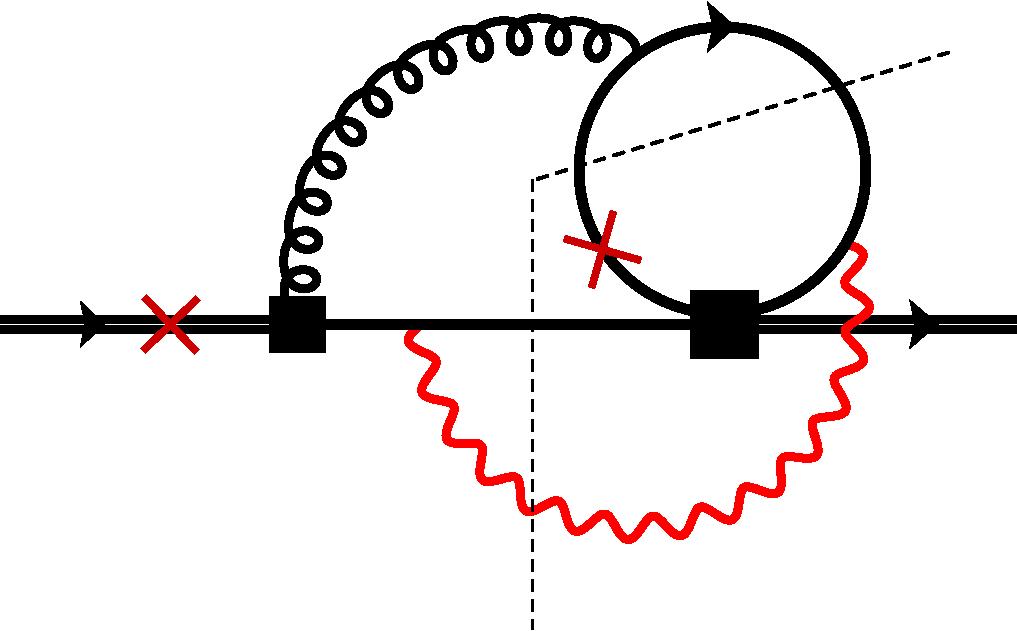}\hspace{5mm}
\raisebox{1.2cm}{\includegraphics[height=1.8cm,width=4.5cm]{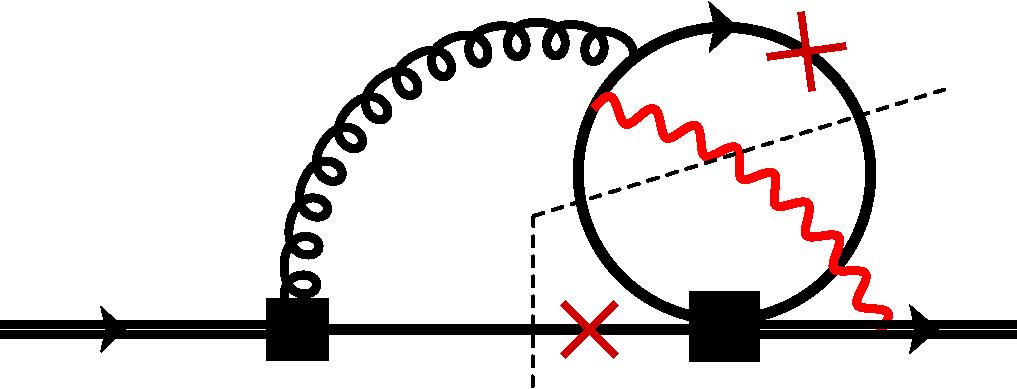}}\\[5mm]
\includegraphics[height=3.4cm,width=4.5cm]{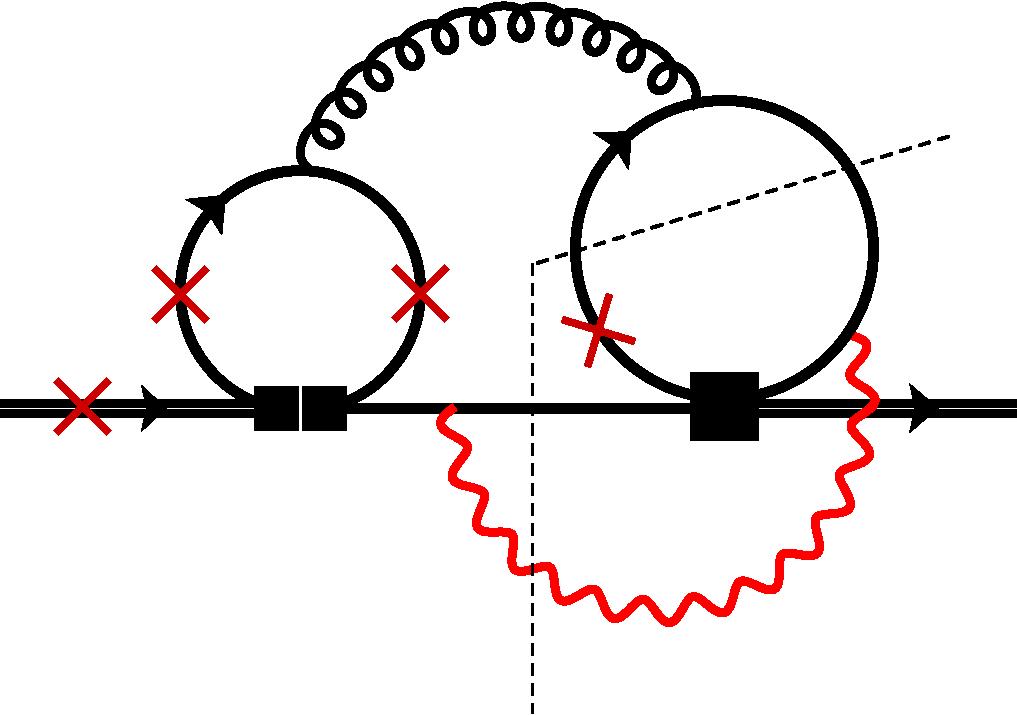}\hspace{5mm}
\raisebox{4.5mm}{\includegraphics[height=2.9cm,width=4.5cm]{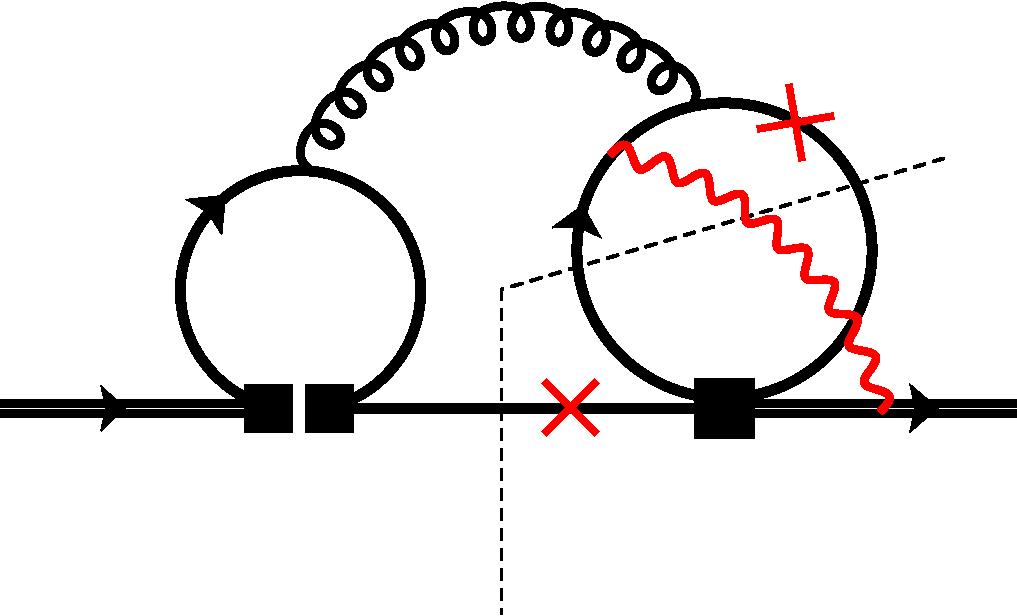}}\hspace{5mm}
\caption{Diagrams of type (ii). Crosses denote alternative insertions of the photon vertex (always one vertex at each side of the cut). Circle-cross denotes the alternative insertion of the gluon vertex.}
\label{diagsii}
\end{figure}

\nind {\bf Type (ii):} The photon couples to the cut fermion loop once (Fig.~\ref{diagsii}): 
In this case the straight diagrams are proportional to $Q_u+Q_d+Q_s=0$ and need not be considered.
The $P_{1,2}^u$ contributions are proportional to $Q_u$.
Therefore the total contribution from these diagrams is:
\eq{{\cal D}^k_{(ii)} = \Big[
Q_d\,\big(-6\,\C3^* + \C4^* - 96\,\C5^* + 16\,\C6^*\big)
+Q_u\,(\C1^{u*}\!-6\,\C2^{u*})
\Big]\av{P_4}^{\times,k}_{(ii)}\ .
\label{eq:(ii)}}\\

\begin{figure}
\centering
\vspace{1cm}
\includegraphics[height=2.3cm,width=5.1cm]{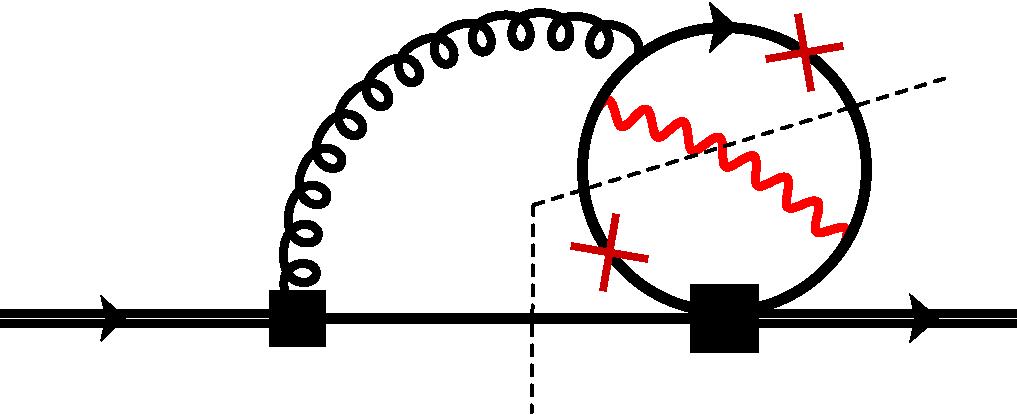}\hspace{1cm}
\raisebox{0.5mm}{\includegraphics[height=2.5cm,width=4.8cm]{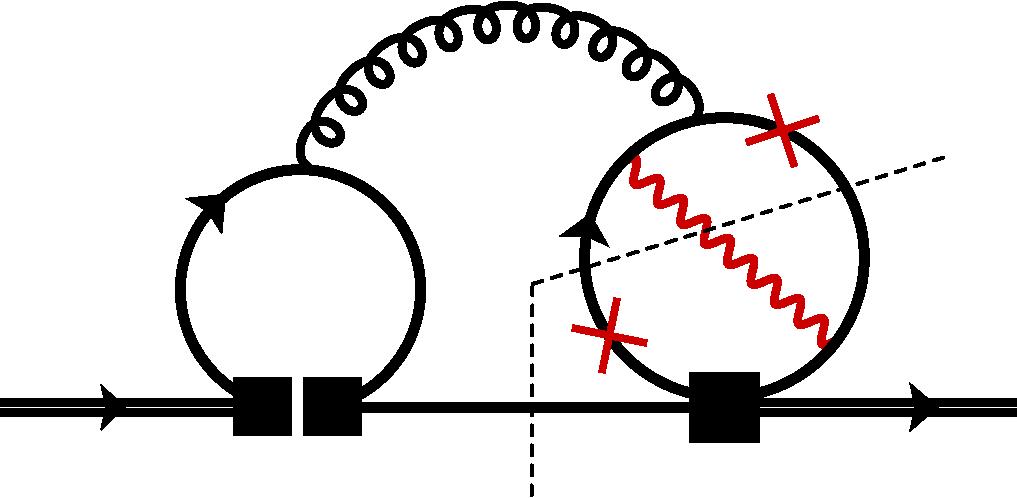}}
\caption{Diagrams of type (iii). Crosses denote alternative insertions of the photon vertex (always one vertex at each side of the cut.}
\label{diagsiii}
\end{figure}

\nind {\bf Type (iii):} The photon couples to the cut fermion loop twice (Fig.~\ref{diagsiii}): In this case crossed
diagrams are proportional to $Q_s^2=Q_d^2$ (or $Q_u^2$ in the case of $P_{1,2}^u$) and straight diagrams to
$Q_u^2+Q_d^2+Q_s^2=Q_u^2+2Q_d^2$. We have:
\eqa{
{\cal D}_{(iii)}^k &=& [Q_u^2+2Q_d^2] \big(\C4^* + 10\,\C6^*\big)\av{P_4}^{s,k}_{(iii)}\nn\\[2mm]
&+&\Big[
Q_d^2\,\big(-6\,\C3^* + \C4^* - 96\,\C5^* + 16\,\C6^*\big)
+Q_u^2\,(\C1^{u*}\!-6\,\C2^{u*})
\Big]\av{P_4}^{\times,k}_{(iii)}\ .
\label{eq:(iii)}}\\
\nind We assume that the objects $\av{P_4}^{I,k}_{(J)}$ are phase-space-integrated matrix elements containing no prefactors
or couplings, such that:
\eq{
\sum_{\begin{minipage}{19mm}$\scriptstyle i=3,..,8, 1q,2q\\[-2mm] j=3,..,6,1u,2u$ \end{minipage}}
\C{i}^*\,\C{j}\,G_{ij}^{(1)} =  
\sum_{J=i,ii,iii} \bigg[ {\cal D}_{(J)}^C +
\sum_{k=3,..,8, 1q,2q} {\cal D}_{(J)}^k \bigg] 
\label{eq:master}}
in the notation of Eq.~(\ref{bsqqgrate}). In Eq.~(\ref{eq:master}), ${\cal D}_{(J)}^C$ denotes the UV counterterms, and both
${\cal D}_{(J)}^C, {\cal D}_{(J)}^k$ include the relevant collinear counterterms.
Both are discussed below in Sections~\ref{sec:renormalization} and \ref{sec:collinear}. The structure
of the objects $\av{P_4}^{(s,\times),k}_{(J)}$ can be deduced from the discussion in Section~\ref{sec:OpId-Left}.
In the case of $P_{7,8}$ we have:
\eqa{
\av{P_4}^{I,7}_{(J)} & = & \C7^\text{eff}\,{\cal F}^{I,7}_{(J)}(\delta)
\hspace{14mm} \text{for}\quad (I,J)=(s,i),(\times,i),(\times,ii)\ ,\hspace{-5mm}\nn\\[4mm]
\av{P_4}^{I,8}_{(J)} & = & \C8^\text{eff}\,Q_d\,\hat{\cal F}^{I,8}_{(J)}(\delta)
\qquad   \text{for}\quad (I,J)=(s,i),(\times,i),(\times,ii)\ ,\hspace{-5mm}\nn\\[4mm]
\av{P_4}^{I,8}_{(J)} & = & \C8^\text{eff}\,\hat{\cal F}^{I,8}_{(J)}(\delta)
\hspace{14mm}  \text{for}\quad I=s,\times \ \ \text{and}\ \  J=iii\ .
\label{eq:P48}
}\\
For $P_{1,2}^{(u)}$ we have:
\eqa{
\sum_{k=1,2} \av{P_4}^{I,k}_{(J)}  &= & (\C1^{c}-6\,\C2^{c})\,\big[Q_d \hat{\cal F}^{I,1}_{(J)}(z_c,\delta)
+ Q_u \widetilde{\cal F}^{I,1}_{(J)}(z_c,\delta) \big]
\nn\\[-1mm]
&&\hspace{49mm}\text{for}\quad (I,J)=(s,i),(\times,i),(\times,ii)\ ,\nn\\[4mm]
\sum_{k=1,2} \av{P_4}^{I,k}_{(J)}  &=& (\C1^{c}-6\,\C2^{c})\,\hat{\cal F}^{I,1}_{(J)}(z_c,\delta)
\qquad \text{for}\quad I=s,\times \ \ \text{and}\ \  J=iii\ ,\qquad
\label{eq:P412}
}
\eqa{
\sum_{k=1u,2u} \av{P_4}^{I,k}_{(J)} &= & (\C1^{u}-6\,\C2^{u})\,\big[Q_d \hat{\cal F}^{I,1}_{(J)}(0,\delta)
+ Q_u \widetilde{\cal F}^{I,1}_{(J)}(0,\delta) \big]
\nn\\[1mm]
&&\hspace{49mm}\text{for}\quad (I,J)=(s,i),(\times,i),(\times,ii)\ ,
\nn\\[4mm]
\sum_{k=1u,2u} \av{P_4}^{I,k}_{(J)}  &=& (\C1^{u}-6\,\C2^{u})\,\hat{\cal F}^{I,1}_{(J)}(0,\delta)
\qquad \text{for}\quad I=s,\times \ \ \text{and}\ \  J=iii\ ,\qquad
\label{eq:P412u}
}
where $z_c \equiv m_c^2/m_b^2$. The contributions with penguin operators to the left of the cut are given by:
\allowdisplaybreaks{
\eqa{
\sum_{k=3...6}\av{P_4}^{I,k}_{(J)}
&= & (\C4-6\,\C3)\,Q_d\,\big[\hat{\cal F}^{I,1}_{(J)}(0,\delta)+\hat{\cal F}^{I,1}_{(J)}(1,\delta) + \widetilde{\cal F}^{I,1}_{(J)}(0,\delta) \big]
\nn\\[1mm]
&+ & 4(\C6-6\,\C5)\,Q_d\,\big[(4-\ep) \big(\hat{\cal F}^{I,1}_{(J)}(0,\delta)+\hat{\cal F}^{I,1}_{(J)}(1,\delta)\big) + 4\,\widetilde{\cal F}^{I,1}_{(J)}(0,\delta) \big]
\nn\\[2mm]
&-& \frac{6\,\C4 + (60-36\ep)\,\C6}{1-\ep}\,Q_d\,
\big[n_\ell\, \hat{\cal F}^{I,1}_{(J)}(0,\delta) + \hat{\cal F}^{I,1}_{(J)}(z_c,\delta) + \hat{\cal F}^{I,1}_{(J)}(1,\delta) \big]
\nn\\[2mm]
&-& 36\,\C6\, Q_u\, \widetilde{\cal F}^{I,1}_{(J)}(z_c,\delta) - 24 (3\,\C5+\C6) Q_d\, \widetilde{\cal F}^{I,1}_{(J)}(1,\delta)
\nn\\[2mm]
&+& (-6\,\C3 +\C4-24\,\C5+4\,\C6)\,Q_d\,\widetilde{\cal F}^{I,4}_{(J)}(\delta)
\nn\\[2mm]
&&\hspace{5.2cm}\text{for}\quad (I,J)=(s,i),(\times,i),(\times,ii)\ ,\nn\\[4mm]
\sum_{k=3...6}\av{P_4}^{I,k}_{(J)}
&=& \big[(\C4-6\,\C3) 
+4(4-\ep)(\C6-6\,\C5)\big]\,\big[\hat{\cal F}^{I,1}_{(J)}(0,\delta) + \hat{\cal F}^{I,1}_{(J)}(1,\delta)\big]
\nn\\
&-& \frac{6\,\C4 + (60-36\ep)\,\C6}{1-\ep}\,
\big[n_\ell\, \hat{\cal F}^{I,1}_{(J)}(0,\delta) + \hat{\cal F}^{I,1}_{(J)}(z_c,\delta) + \hat{\cal F}^{I,1}_{(J)}(1,\delta) \big]
\nn\\[2mm]
&&\hspace{5.2cm}\text{for}\quad I=s,\times \ \ \text{and}\ \  J=iii\ .\qquad
\label{eq:P43}
}
}

The functions $\hat{\cal F}^{I,k}_{(J)}={\cal F}^{I,k}_{(J)}+{\cal F}^{I,k}_{\text{coll}(J)}$ include the collinear
regulators discussed in Section~\ref{sec:collinear}. The functions $\widetilde{\cal F}^{I,1}_{(J)}$ and $\widetilde{\cal F}^{I,4}_{(J)}$ are related to diagrams
where the photon couples to the left-hand quark loop, corresponding respectively to the terms with $h^{\mu\nu}$ and $\widetilde h^{\mu\nu}$ in Eq.~(\ref{eq:3charm}).
Explicit results for all these functions are collected in Appendix~\ref{sec:interresults}.

\subsection{Other details}

\subsubsection{Irrelevance of evanescent terms to the right of the cut}

In the case of $(P_7,P_i)$ interference, there are no UV or collinear divergences, and therefore evanescent structures are irrelevant for the $\ord{\ep^0}$ result.

In the case of $(P_8,P_i)$ interference, collinear divergences appear which combined with evanescent terms give finite
contributions in the dimensionally regularized result. However these finite terms cancel when we express the dimensional
regulators in terms of logarithms of masses, via the splitting-function approach (see Section~\ref{sec:collinear}):
\eq{\frac{d\Gamma}{dx} = \frac{d\Gamma_\ep}{dx} + \frac{d\Gamma_{\rm shift}}{dx} + \ep \bigg( \frac{d\Gamma_\ep^{\rm Ev}}{dx} + \frac{d\Gamma_{\rm shift}^{\rm Ev}}{dx}\bigg) = \frac{d\Gamma_{\rm mass\ reg.}}{dx} +\ord\ep
}
since the $1/\ep$ terms cancel in both 4D and evanescent terms separately.

In the case of $(P_{1,2},P_i)$ interference, UV and collinear divergences are nested inside dimensionally regularized expressions. However all UV divergences cancel against counterterm diagrams, including finite terms from evanescent operators:
\eq{
\hspace{-10mm}
\raisebox{-13mm}{
\includegraphics[width=8.5cm]{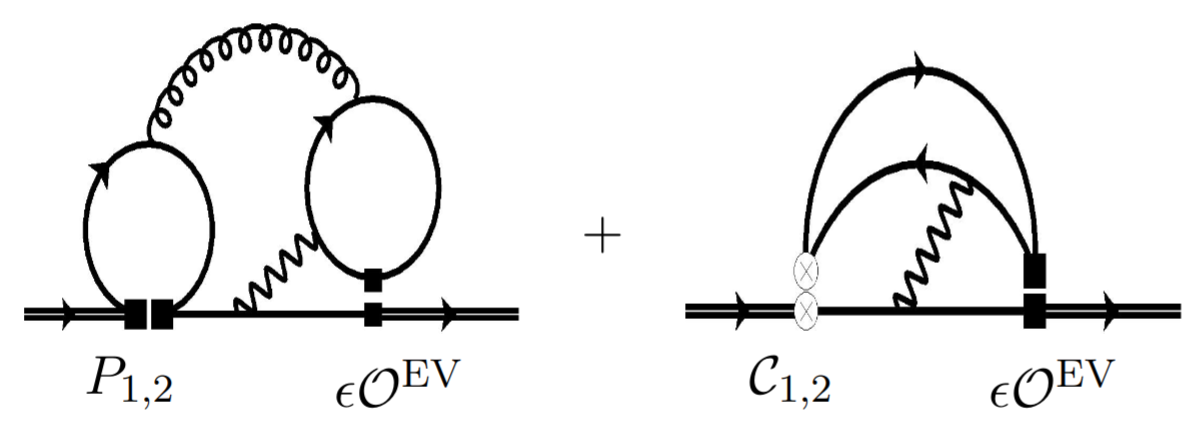}}
=\quad \ep\ \bigg(\frac1{\ep_{\rm coll}} + \text{UV finite}\bigg)\hspace{-0.5cm}
\label{NoEv}
}
All ``finite'' terms from collinear divergences now disappear when going to mass regularization, as in the case with $P_8$.

\subsubsection{Cancellation of $i \epsilon_{\mu\nu\rho\sigma} k_1^\mu k_2^\nu k_3^\rho k_4^\sigma$ terms}
\label{sec:epsterms}

Traces with $\g_5$ will introduce terms proportional to $i \epsilon_{\mu\nu\rho\sigma} k_1^\mu k_2^\nu k_3^\rho k_4^\sigma$ in the differential decay rate. Here we show that these terms always cancel if we perform a full angular integration over phase-space.

Consider fixing all double invariants $k_i\cdot k_j$. Then all $k_i$ are fixed only up to an Euler rotation \emph{and} an orientation. To see this go to the rest frame of the $b$-quark. Momentum conservation fixes all the energies (since $k_i\cdot p_b$ are fixed). This implies that $\vec k_i\cdot \vec k_j$ are also fixed. We can rotate the frame to put $\vec k_1$ along the positive $z$ axis, and $\vec k_2$ in the $(y,z)$ plane. Then $\vec k_3$ is fixed only up to a two-fold ambiguity (an orientation), given by the sign of its $x$ component. Once this sign is chosen $k_4$ is also fixed. This proves that $i \epsilon_{\mu\nu\rho\sigma} k_1^\mu k_2^\nu k_3^\rho k_4^\sigma$ is fixed by $k_i\cdot k_j$ up to a sign, which is given by the orientation of $(\vec k_1,\vec k_2,\vec k_3)$.

Now consider phase-space integration. Terms in the integrand of the form $F(k_i\cdot k_j)$ do  not depend on the Euler rotation nor the orientation, and the angular integral over $d\Omega_{3} d\Omega_{2}d\Omega_{1}$ can always be performed trivially, giving a factor $16\pi^2$. Terms of the form $F(k_i\cdot k_j)\epsilon_{\mu\nu\rho\sigma} k_1^\mu k_2^\nu k_3^\rho k_4^\sigma$, however, change sign under change of orientation, and vanish upon integration over $d\Omega_1$. Obviously parity-odd terms cancel out in parity-even observables. Therefore we drop these terms from the beginning in the calculation of the integrated decay rate.\\

\subsection{Phase-space integration}
\label{sec:PSint}

The phase-space measure for a ($1\to 4$) decay of a particle of mass $M$ into four massless particles with momenta $k_{1,2,3,4}$ is given in terms of kinematic invariants by \cite{0311276}:
\eqa{
dPS_4 &=& \tilde \mu^{6\ep}\, 2^{5-5d} \pi^{4-3d} M^{3d-8}\, (-\Delta_4)^{\frac{d-5}2}\,
\delta(1-s_{12}-s_{13}-s_{14}-s_{23}-s_{24}-s_{34})\nn\\
&&\times\,\Theta(-\Delta_4)\, d\Omega_{d-1}\, d\Omega_{d-2}\, d\Omega_{d-3}\, ds_{12}\, ds_{13}\, ds_{14}\, ds_{23}\, ds_{24}\, ds_{34}\ ,
}
where $s_{ij}=2k_i\cdot k_j/M^2$ ($0 \leq s_{ij} \leq 1$), and $\Delta_4$ is the Gram determinant:
\eq{+\Delta_4 = s_{12}^2 s_{34}^2 + s_{13}^2 s_{24}^2 + s_{14}^2 s_{23}^2 -2 s_{12} s_{34} s_{13} s_{24} -2 s_{12} s_{34} s_{14} s_{23} -2 s_{13} s_{24} s_{14} s_{23}\ .}
The unpolarized decay rate is given by the phase-space integral:
\eq{\Gamma = \frac1{2M}\frac{1}{2N_c}\int \sum |{\cal M}|^2 dPS_4}
where the sum runs over the spins and color of all particles (we assume the parent is a color triplet). $\sum |{\cal M}|^2$ depends only on $s_{ij}$: $\sum |{\cal M}|^2\equiv {\cal K}(s_{ij})$, so the angular integrations can be performed trivially:
\eq{\int d\Omega_{d-1}\, d\Omega_{d-2}\, d\Omega_{d-3} = \frac{8 \pi^{\frac{3d-6}2}}{\Gamma(\frac{d-1}2)\Gamma(\frac{d-2}2)\Gamma(\frac{d-3}2) }\ ,}
and the general formula for the decay rate becomes
\eq{
\Gamma = \frac{\tilde \mu^{6\ep}\,2^{8-5d}\,\pi^{1-3d/2} M^{3d-9}}{4N_c \Gamma(\frac{d-1}2)\Gamma(\frac{d-2}2)\Gamma(\frac{d-3}2)}\int [ds_{ij}]\,\delta(1-{\textstyle \sum} s_{ij}) {\cal K}(s_{ij})  (-\Delta_4)^{\frac{d-5}2}\,\Theta(-\Delta_4)\ .
\label{genPS}}
This integral might contain soft and/or collinear divergences associated to regions of phase space where some particles are soft or collinear. These divergences can be regularized in dimensional regularization by setting $d=4-2\epsilon$. If we insist on integrating over these regions, one must include virtual corrections to cancel the divergences. Otherwise, the regulator must be traded by a physical cutoff at a later stage.\\

We now specify to the $b\to q(k_1)\bar q(k_2) s(k_3) \gamma(k_4)$ case. We consider a cut on the photon energy $E_\gamma > E_0 \equiv \frac{m_b}{2}(1-\delta)$ (in the $b$ quark rest frame), which defines the parameter $\delta$. This translates into the constraint $s_{14}+s_{24}+s_{34}>1-\delta$, which can be included in the phase-space integral in the following way. We include a delta function $\delta(1-z-s_{14}-s_{24}-s_{34})$ in the integrand, and we integrate over the new variable $z$ from $0$ to $\delta$:
\eq{
\int_0^\delta dz\int_0^1 [ds_{ij}]\,\delta(1-z-s_{14}-s_{24}-s_{34})\delta(z-s_{12}-s_{23}-s_{13})  {\cal K}(s_{ij}) 
 (-\Delta_4)^{\frac{d-5}2}\,\Theta(-\Delta_4)\ .
}
The delta functions can be used to integrate over two invariants, e.g.\ $s_{13}$ and $s_{24}$:
 \eq{
 \Gamma_{E_\gamma>E_0} = N(d)\int_0^\delta\!\! dz \int_0^{\bar z}\!\! ds_{34}\int_0^{\bar z-s_{34}}\hspace{-8mm} ds_{14}
 \int_0^{z}\!\! ds_{12}\int_0^{z-s_{12}}\hspace{-8mm} ds_{23}\ {\cal K}(s_{ij})  (-\Delta_4)^{\frac{d-5}2}\,\Theta(-\Delta_4)
 \Big|_{s_{13},s_{24}}
 }
where $\bar z\equiv 1-z$, and $N(d)$ is given by the prefactor in Eq.~(\ref{genPS}), and the substitution rule $X|_{s_{13},s_{24}}$ corresponds to $s_{13} \to z-s_{12}-s_{23}$ and $s_{24} \to \bar z-s_{14}-s_{34}$.
The next integration can be performed over an invariant that appears only polynomially 
in ${\cal K}$ (see e.g.~\cite{0512066}). It is easy to see that $s_{23}$ always satisfies 
this criterion by checking the uncut propagators in Figs.~\ref{diagsi},\ref{diagsii},\ref{diagsiii} 
and the loop functions. Upon substitution of $s_{13},s_{24}$, the Gram determinant remains quadratic 
in $s_{23}$: $-\Delta_4=(\bar z-s_{34})^2 (a^+-s_{23})(s_{23}-a^-)$, where $a^\pm$ 
are complicated functions of the rest of the invariants:
\eqa{
(\bar z-s_{34})^2 a^\pm &=& (\bar z -s_{34}) [{\vtwo{z(\bar z-s_{34}-s_{14})-s_{12}(\bar z +s_{14})}}]
{\vtwo{+2 \, \bar z \, s_{12} \, s_{14}}}
\pm 2 \sqrt{\Xi}\ ,\qquad\\[2mm]
\Xi &=& s_{12} s_{14} s_{34} (s_{14} + s_{34} - \bar z) [z s_{34} - \bar z (z - s_{12})]\ .
}
Thus, $-\Delta_4$ is positive only if $a^\pm$ are real (happening only if $s_{34}<\bar z(z - s_{12})/z<\bar z$), and for $a^-<s_{23}<a^+$. In addition, $a^->0$. This sets the integration limits for $s_{23}$ and $s_{34}$ imposed by the $\Theta$-function, which can then be dropped:
\eq{
\Gamma_{E_\gamma>E_0} = N(d)\int_0^\delta\!\! dz \int_0^{z}\!\! ds_{12}\int_0^{\bar z(z-s_{12})/z}
\hspace{-10mm} ds_{34} \int_0^{\bar z-s_{34}}\hspace{-8mm} ds_{14} \int_{a^-}^{a^+}\hspace{-3mm} ds_{23}
\ {\cal K}(s_{ij})(-\Delta_4)^{\frac{d-5}{2}}
\Big|_{s_{13},s_{24}}\ .
}
Now it is convenient to perform the following changes of variables:
\eq{
\begin{array}{rclrcl}
s_{12} &=& z v w & s_{34} &=& \bar z \bar v \\[2mm]
s_{14} &=& \bar z v x \hspace{15mm} & s_{23} &=& (a^+-a^-) u + a^- 
\end{array}
\label{cvars}}
where $u,v,w,x$ are integrated independently from 0 to 1, and
\eqa{
(a^+-a^-) &=& 4 z (\bar v w \bar w x \bar x)^{1/2}\ , \label{apmam}\\
a^-\quad &=& z \big[ \bar v w x +  \bar w \bar x -2 (\bar v w \bar w x \bar x)^{{\vtwo{}}1/2} \big]\ .\label{am}
}
This gives
\eqa{
\Gamma_{E_\gamma>E_0} &=& N(d)\int_0^\delta\!\! dz\,z\,\bar z^{d-3} \int_0^{1}\!\!du\,dv\,dw\,dx\,
(u \bar u)^{\frac{d-5}{2}} v^{d-3} (a^+-a^-)^{d-4}
\ {\cal K}\nn\\
&= & N(d)\,4^{d-4}\int_0^\delta\!\! dz\,(z\bar z)^{d-3} \int_0^{1}\!\!du\,dv\,dw\,dx\,
 (u\bar u)^{\frac{d-5}{2}}
 v^{d-3} (\bar v w\bar w x \bar x)^\frac{d-4}{2} \ {\cal K}\ .\qquad
}
%
In the following we must consider the kernel ${\cal K}(u,v,w,x,z)$. 
As mentioned above, ${\cal K}$ is polynomial in $s_{23}$: ${\cal K} = \sum f_n(v,w,x,z) s_{23}^n$. 
Expanding $s_{23}$ according to (\ref{cvars})-(\ref{am}) will provide a sum of terms of the form
\eq{
{\cal K} = \sum_{m,n} f_{n,m}(v,w,x,z)\ u^m\ ,
}
and the integral over the variable $u$ gives a factor $\beta\big(\frac{d-3}2+m,\frac{d-3}2\big)$ for each term.
The next steps depend on the diagram at hand. Consider the diagrams with $P_{7,8}$. In this case,
\eq{
f_{n,m}(v,w,x,z) = v^a\, \bar v^b\, w^c\, \bar w^e\, x^f\, \bar x^g\, z^h\, \bar z^p\, (1-\bar z \bar v)^q\, (1- z \bar w)^r
}
for some $a,b,c,... \in R$. The integral over $x$ gives again a $\beta$-function: $\beta\big(\frac{d-2}2+f,\frac{d-2}2+g\big)$.
Because of the $(1-\bar z \bar v)^q$ and $(1-z \bar w)^r$ factors, the next steps will introduce hypergeometric functions.
The integral over $v$ gives
\eq{
\beta\left(d-2+a,\frac{d-2}2+e\right) {}_2F_1\left(-q,\frac{d-2}2+b;\frac{3d-6}2+a+b;\bar z\right)\ ,
}
and the integral over $w$, gives:
\eq{
\beta\left(\frac{d-2}2+c,\frac{d-2}2+e\right) {}_2F_1\left(-r,\frac{d-2}2+e;d-2+c+e;z\right)\ .
}
The next step is to expand around $\ep\to 0$ (with $d=4-2\ep$).
The expansion of hypergeometric functions is performed automatically by the package {\tt HypExp} \cite{HypExp}.
This will give finite results in the case of $P_7$, but $1/\ep$ poles in the case of $P_8$,
corresponding to collinear divergences. The integration over the photon energy $z\in (0,\delta)$ can then be
performed, also analytically, for all terms.

The case of loop diagrams is in principle more complicated, as the function ${\cal K}$ contains already
a hypergeometric function. For instance, in the case of diagrams~(i) with the photon {\it not} attaching to the
quark loop, the variable $s_{12}$ appears in the function $g(m_q)\sim {}_2F_1(\ep,2;\frac52;\frac{s_{12}}{4 z_q})$
(cf.\ Eq.~(\ref{eq:g(m)})). However, 
by a suitable choice in the order of integration, analytic results can be obtained as before. In the case of
diagrams such as (iii), the hypergeometric function depends on the triple invariant $s_{124}=s_{12}+s_{14}+s_{24}$, and the
sequential-integration procedure described above does not seem to work up to finite order in $\epsilon$. In this case we
extract all the $1/\ep^2$ and $1/\ep$ poles analytically and leave the finite terms differential in one of the
variables, which we integrate numerically afterwards. This is also the case for the diagrams where the photon
couples to the charm loop, which are both UV and collinear finite. In general, for the loop contributions, some finite
terms turn out to be complicated functions of $\delta$ and $z_c \equiv m_c^2/m_b^2$. We give these results as polynomial
expansions in $\delta$ around the physical value $\delta=0.316$. The coefficients of this expansion are presented as numerical
interpolations in the variable $z_c$, reproducing the exact results to enough precision for all practical purposes.
We have checked that the interpolated expressions in the appendix reproduce the exact results with high precision
in the full range $z_c\in (0,1)$ for values of $\delta$ near $0.316$.

\subsection{Renormalization}
\label{sec:renormalization}

Tree-level four-body contributions from four-quark operators arise at LO in $\alpha_s$ and have been computed in Ref.~\cite{1209.0965}.
At NLO the corresponding counterterm contributions must be included, which cancel the UV divergences from the loop diagrams.
One must consider the insertion of the bare operators $P_i^{(0)}$, $i=1q,2q,3,..,6$, in the tree-level diagrams to the left of the cut, where:
\eqa{
\sum_{i=3,..,6, 1q,2q} \C{i}\, P_i^{(0)}
&=& \sum_{\begin{minipage}{18mm}$\scriptstyle i=3,..,6, 1q,2q\\[-2mm] j=3,..,6,1u,2u$ \end{minipage}} \C{i}\, Z_{ij} P_j
= \sum_{\begin{minipage}{18mm}$\scriptstyle i=3,..,6, 1q,2q\\[-2mm] j=3,..,6,1u,2u$ \end{minipage}}
\C{i}\, \bigg(\delta_{ij} + \frac1{\ep}\frac{\alpha_s}{4\pi} \delta Z_{ij}\bigg) P_j\nn\\[2mm]
&=& \sum_{i=3,..,6, 1u,2u} \C{i}\, P_i  + \frac1{\ep}\frac{\alpha_s}{4\pi}
\sum_{\begin{minipage}{18mm}$\scriptstyle i=3,..,6, 1q,2q\\[-2mm] j=3,..,6,1u,2u$ \end{minipage}} \C{i}\, \delta Z_{ij} P_j
}
The first term leads to the LO contributions in Ref.~\cite{1209.0965}, while the second term contributes to the NLO result
and takes care of the UV divergences. For this we need, a priori, the tree level results with $P_{3,..,6,1u,2u}$ including $\ord{\ep}$
terms, and the renormalization factors $\delta Z_{ij}$.

\begin{figure}
\centering
\includegraphics[width=15cm]{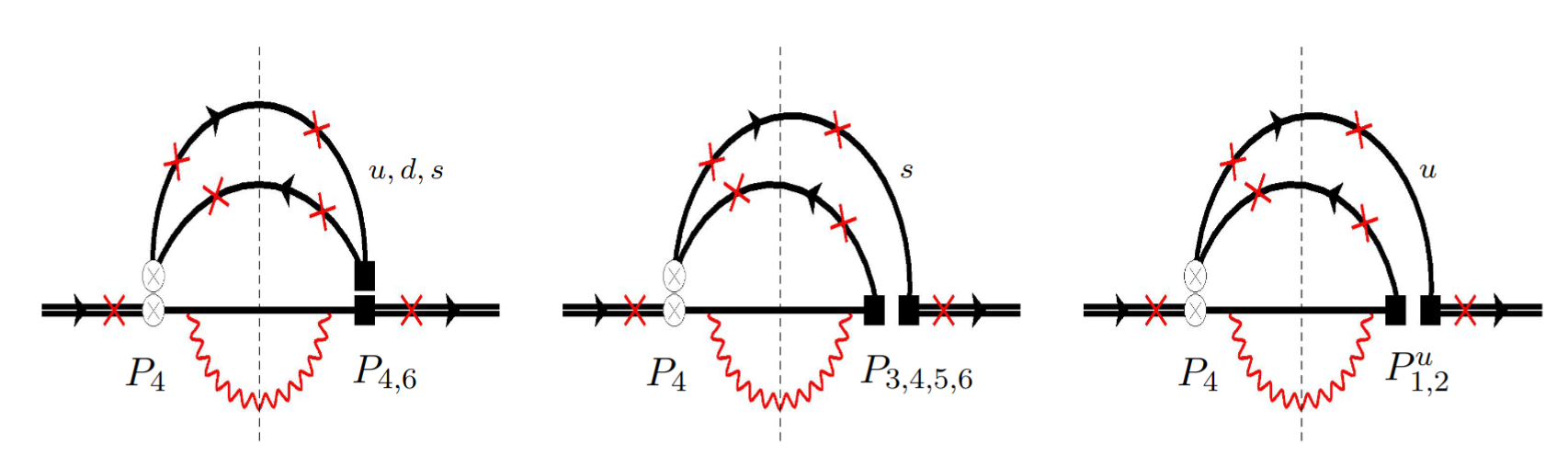}
\vspace{-2mm}
\caption{Tree-level counterterm diagrams. Crosses denote alternative insertions of the photon vertex (always one vertex at each side
of the cut). These diagrams can be classified in types (i), (ii), (iii) as done for the loop diagrams.}
\label{fig:counterterms}
\end{figure}

The relevant renormalization factors are simple to compute. Using the relationships developed in Section~\ref{sec:OpId},
and expressing the result in terms of tree-level matrix elements of four-quark operators, we find that:

\eq{
\raisebox{-4mm}{\includegraphics[height=1.8cm,width=2.2cm]{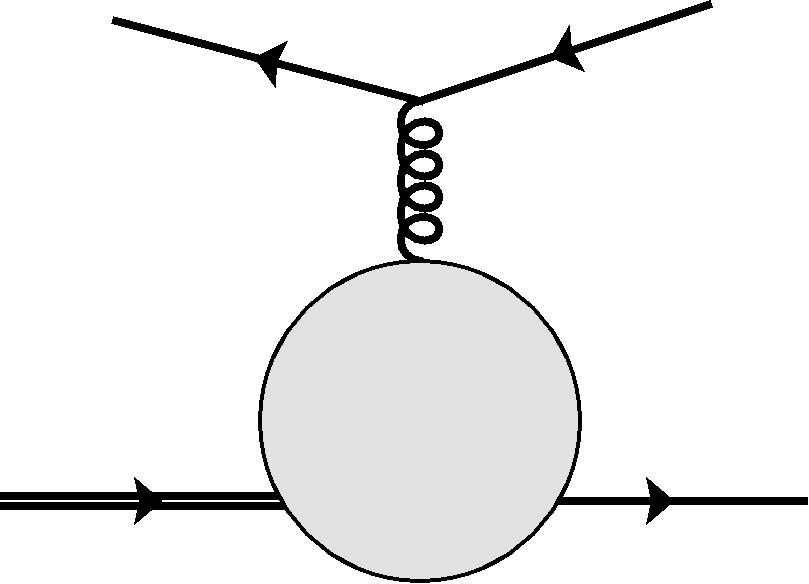}}
\hspace{5mm}
\begin{array}{cl}
=\quad & \bigg[ \C1^u-6\,\C2^u+\C1^c-6\,\C2^c-12\,\C3 -28\,\C4-192\,\C5 -268\,\C6 \bigg] \\[5mm]
&\displaystyle \times\ \frac19 \frac1{\ep}\frac{\alpha_s}{4\pi} \av{P_4}^\text{tree} + \ord{\ep^0}\ .
\end{array}
}
This fixes the renormalization factors needed in our calculation:
\eq{
\begin{array}{llll}
\delta Z_{1u\,4} = -\frac19 \qquad& \delta Z_{1c\,4} = -\frac19 \qquad& \delta Z_{34} = \frac43    \qquad& \delta Z_{44} = \frac{28}9 \\[3mm]
\delta Z_{2u\,4} =  \frac23 \qquad& \delta Z_{2c\,4} =  \frac23 \qquad& \delta Z_{54} = \frac{64}3 \qquad& \delta Z_{64} = \frac{268}9\ .
\end{array}
\label{eq:Zfactors}}
We also see that we need only tree level diagrams with insertion of $P_4$ to the left of the cut. All the diagrams needed
are shown in Fig.~\ref{fig:counterterms}.

For the operator insertions to the right of the cut we can (and must) use the 4D identities derived in Section~\ref{sec:OpId},
noting that evanescent terms cancel in the renormalization process by virtue of Eq.~(\ref{NoEv}). This leads to exactly the
same structure as Eqs.~(\ref{eq:(i)}),(\ref{eq:(ii)}),(\ref{eq:(iii)}) for the counterterm diagrams ${\cal D}_{(J)}^{C}$
(i.e.\ Eq.~(\ref{eq:master})), with the corresponding matrix elements $\av{P_4}_{(J)}^{I,C}$ given by:
\eqa{
\av{P_4}^{I,C}_{(J)} &=& \bigg[\C1^u-6\,\C2^u+\C1^c-6\,\C2^c-12\,\C3 -28\,\C4-192\,\C5 -268\,\C6\bigg]\nn\\[2mm]
&& \times \left\{
\begin{array}{ll}
Q_d \hat{\cal F}^{I,C}_{(J)}(\delta) & \quad \text{for}\quad (I,J)=(s,i),(\times,i),(\times,ii)\\[3mm]
\hat{\cal F}^{I,C}_{(J)}(\delta) &  \quad\text{for}\quad I=s,\times \ \ \text{and}\ \  J=iii\ .
\end{array}
\right.
\label{eq:P4C}
}
Again, $\hat {\cal F}^{I,C}_{(J)} = {\cal F}^{I,C}_{(J)} + {\cal F}^{I,C}_{\text{coll}(J)}$. The functions
${\cal F}^{I,C}_{(J)}, {\cal F}^{I,C}_{\text{coll}(J)}$ are given in Appendix~\ref{sec:interresults}. One can check that all UV
divergences cancel, as expected: $\av{P_4}^{I,C}_{(J)} + \sum_k \av{P_4}^{I,k}_{(J)} = \text{UV finite}$.

\subsection{Collinear divergences and splitting functions}
\label{sec:collinear}

The region of phase space in which the photon is collinear to one of the light quarks gives rise to collinear divergences.
These divergences are regulated dimensionally in our computation. However, these are just artifacts of the massless limit
used for light quarks, and there is a more natural regulator: a physical cut-off given by the light meson masses. A suitable
parametrization of such (near-) collinear effects consists in keeping the light quarks massive and perform a massive
phase-space integration. This is quite complicated from the practical point of view, taking into account that the massless
phase-space integrals computed here are already rather challenging.

\begin{figure}
\centering
\includegraphics[height=2.6cm,width=4.3cm]{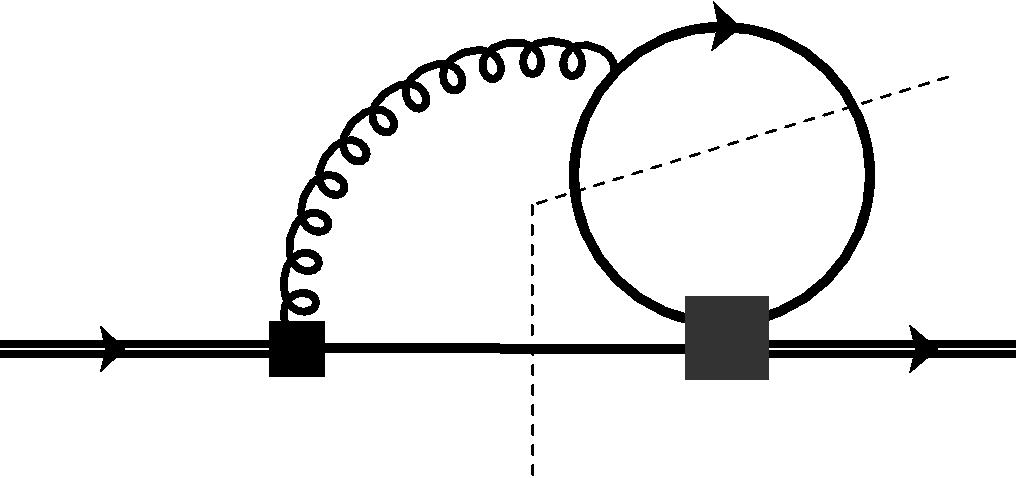}\hspace{8mm}
\includegraphics[height=3.2cm,width=4.3cm]{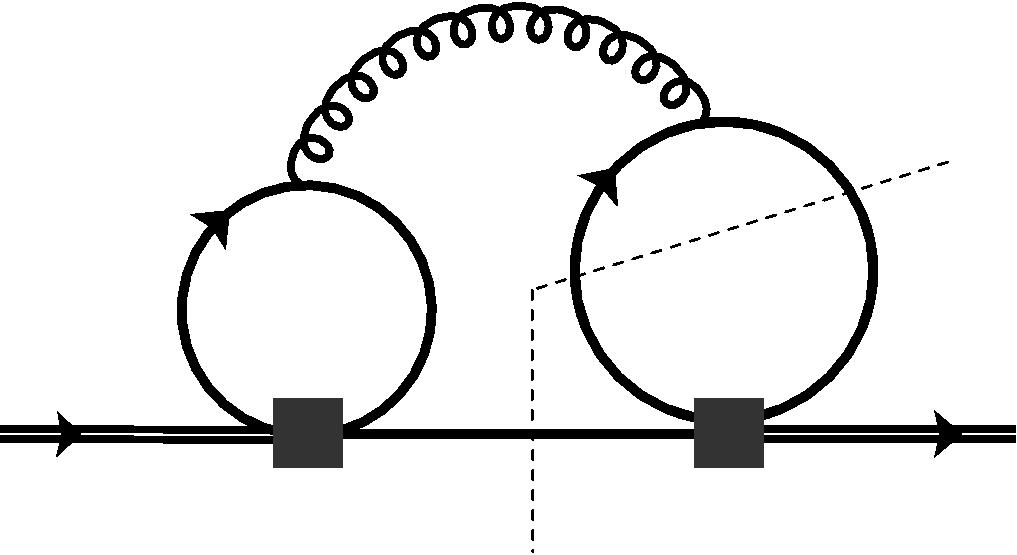}\hspace{8mm}
\includegraphics[height=2.8cm,width=4.3cm]{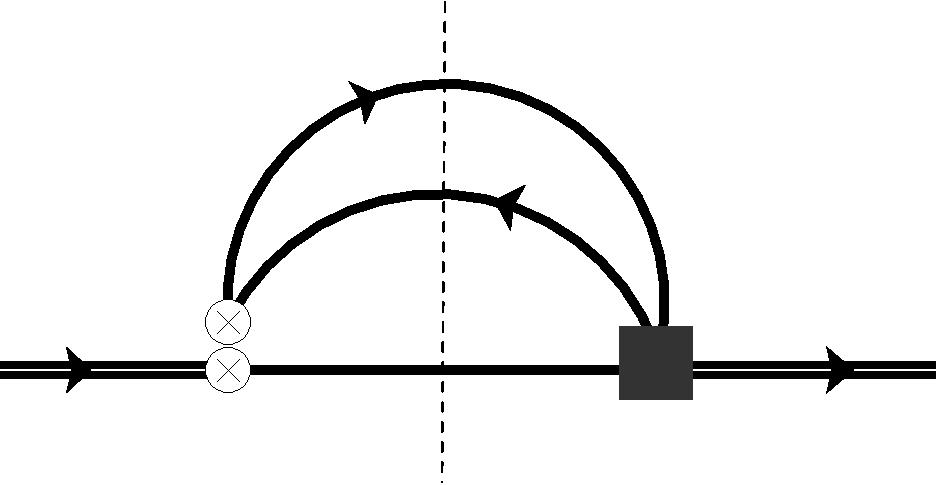}
\caption{Three-particle-cut diagrams needed for the calculation of collinear terms.}
\label{fig:collinear}
\end{figure}

Fortunately, one may resort to the factorization properties of the amplitudes in the quasi-collinear limit (see e.g.\
\cite{1209.0965}).
The idea is that close to the collinear region, the $b\to q_1 \bar q_2 q_3 \gamma$ amplitude may be expressed as a
$b\to q_1 \bar q_2 q_3$ amplitude times a \emph{splitting function} $f_i$, describing the quasi-collinear emission of a
photon from $q_i$, summed over $i=1,2,3$. The splitting functions encode the collinear divergences, and can themselves be
regulated by quark masses or in dimensional regularization. Both approaches are rather simple, since in this limit the four-body
phase space factorizes into a convolution of the three-body phase space of the non-radiative process and the phase space
of the radiative process alone: $d\Phi_4=d\Phi_3\otimes d\Phi$. By comparing the splitting functions regulated in these
two different schemes, one can write a formula to switch from one to the other at the level of the decay rate
\cite{1209.0965}:
\eq{
\frac{d\Gamma_\text{m}}{dz} = \frac{d\Gamma_\ep}{dz} + \frac{d\Gamma_\text{shift}}{dz}
}
where
\eqa{
 \frac{d\Gamma_\text{shift}}{dz} &=& \frac1{2 m_b}\frac1{2N_c} \int dPS_3\, {\cal K}_3(s_{ij})\,
\frac{\alpha_e}{2\pi \bar z}
\left\{
Q_1^2 \bigg[ 1+\frac{(z - s_{23})^2}{(1-s_{23})^2} \bigg] \right. \nn\\[3mm]
&&
\left. \times \bigg[ \frac1\ep -1 + 2\log \frac{(1-s_{23})\mu}{m_{q_1} (1-z)} \bigg] \Theta(z-s_{23}) + (\text{cyclic})
\right\}\ .
\label{Gshift}
}
Here ${\cal K}_3=\sum |{\cal M}_3|^2$ is the spin-summed squared matrix element of the $b\to q_1 \bar q_2 q_3$ decay
obtained by evaluating the diagrams in Fig.~\ref{fig:collinear}, and $dPS_3$ is the three-particle phase-space measure in
$d=4-2\ep$ dimensions \cite{0311276}:
\eqa{
dPS_3 &=& \tilde \mu^{4\ep}\, 2^{2-3d} \pi^{3-2d} m_b^{2d-6}\, (s_{12} s_{13} s_{23})^{\frac{d-4}2}\,\delta(1-s_{12}-s_{13}-s_{23})\nn\\
&&\times\, d\Omega_{d-1} d\Omega_{d-2}\, ds_{12} ds_{13} ds_{23}\ .
}

Integrating Eq.~(\ref{Gshift}) over $z \in [0,\delta]$ provides the terms ${\cal F}_\text{coll}$ contained in the functions~$\hat {\cal F}$.
The contributions from the chromomagnetic operator (Fig.~\ref{fig:collinear}, left) enter into Eq.~(\ref{eq:P48}).
The contributions from four-quark operators (Fig.~\ref{fig:collinear}, center) go into
Eqs.~(\ref{eq:P412}), (\ref{eq:P412u}) and (\ref{eq:P43}). The counterterm contributions
(Fig.~\ref{fig:collinear}, right), enter into Eq.~(\ref{eq:P4C}). The functions~${\cal F}_\text{coll}(\delta)$ are
collected in Appendix~\ref{sec:interresults}.

One can check that adding the collinear contribution removes the $1/\ep$ terms
that survive the renormalization process, trading them for collinear logarithms of quark-mass ratios.
These collinear logarithms are of the form $\log(m_q/m_b)$, with $q=u,d,s$.
The quark masses are collinear regulators and it is difficult to relate them to physical masses.
In our numerical analysis we will take a common constituent-quark mass $m_q\sim 100-250$ MeV for all three light flavors,
and use the notation $L_q = \log(m_q/m_b) \sim \log(m_u/m_b)\sim \log(m_d/m_b)\sim \log(m_s/m_b)$.
This should provide a reasonable estimate of the effect of collinear logarithms.

\section{Results}
\label{sec:results}

We write the four-body contribution to the $\bar B\to X_s\gamma$ rate as:
\eq{
\Delta \Gamma (\bar B\to X_s\gamma)_{E_\gamma>E_0}^{s\bar q q\gamma}
= \Gamma_0 \sum_{i,j} \C{i}^\text{eff}(\mu_b)^*\,\C{j}^\text{eff}(\mu_b)\,G_{ij}(\mu_b,\delta)\ ,
\label{eq:DeltaGamma}}
where $\Gamma_0$ is the absolute normalization of the decay rate:
\eq{
\Gamma_0 = \frac{G_F^2 \alpha_e m_b^5 |V_{ts}^* V_{tb}|^2}{32\pi^4}\ .
}
The sum runs over $i,j = 1u,2u,3,..,6,1c,2c,7,8$. The Wilson coefficients $\C{3,..,8}$ are real, but $\C{1u,2u,1c,2c}$ contain
CKM phases:
\eqa{
\C{3,..,6}^\text{eff} &=& C_{3,..,6}\ , \quad \C{1u,2u}^\text{eff} = -\frac{V_{us}^*V_{ub}}{V_{ts}^*V_{tb}} C_{1,2}
\ , \quad \C{1c,2c}^\text{eff} = -\frac{V_{cs}^*V_{cb}}{V_{ts}^*V_{tb}} C_{1,2}\ , \nn\\
\C7^\text{eff} &=& C_7- \frac13 C_3 - \frac{4}{{\vtwo{9}}} C_4 - \frac{20}3 C_5 - \frac{80}9 C_6\ ,\nn\\
\C8^\text{eff} &=& C_8+C_3 - \frac16 C_4 + 20 C_5 - \frac{10}3 C_6\ ,
}
with $C_i$ the Wilson coefficients in the notation of Ref.~\cite{9612313}. They are needed here to NLO:
\eq{
\C{i}^\text{eff}(\mu) = \C{i}^\text{(0)eff}(\mu) + \frac{\alpha_s(\mu)}{4\pi} \C{i}^\text{(1)eff}(\mu) + \ord{\alpha_s^2},
}
their numerical values are given below.

The matrix elements $G_{ij}(\mu,\delta)$ depend on the renormalization scale and the photon-energy cut and can be split into
LO and NLO components:
\eq{
G_{ij}(\mu,\delta) = G_{ij}^{(0)}(\delta) + \frac{\alpha_s(\mu)}{4\pi} G_{ij}^{(1)}(\mu,\delta) + \ord{\alpha_s^2}\ .
}
The LO matrix $G^{(0)}$ is real and symmetric and was computed in Ref.~\cite{1209.0965}:
we reproduce and confirm these results (after the 2014 update of that paper).
We write (here $i,j=1u,2u,3,..,6$):
\allowdisplaybreaks{
\eq{
\small
\hspace{-2mm} G^{(0)}(\delta)= \left(
\begin{array}{cccccc}
\frac29 T_2 &  0 & \frac49 T_2 &  -\frac2{27} T_2 &  \frac{64}9 T_2 &  -\frac{32}{27}T_2 \\[1mm]
0 &  T_2 &  \frac13 T_2 &  \frac49 T_2 &  \frac{16}3 T_2 &  \frac{64}9 T_2 \\[1mm]
\frac49 T_2 &  \frac13 T_2 &  T_1 + T_3 &  \frac43 T_3 &  10T_1 + 16T_3 &  \frac{64}3 T_3 \\[1mm]
-\frac2{27} T_2 &  \frac49 T_2 &  \frac43 T_3 &  \frac29 (T_1 - T_3) & \frac{64}3T_3 &  \frac{20}9T_1 - \frac{32}9T_3 \\[1mm]
\frac{64}9 T_2 &  \frac{16}3 T_2 &  10T_1 + 16T_3 &  \frac{64}3 T_3 &  136T_1 + 256T_3 &  \frac{1024}3 T_3 \\[1mm]
-\frac{32}{27} T_2 &  \frac{64}9 T_2 &  \frac{64}3 T_3 &  \frac{20}9 T_1 - \frac{32}9 T_3 &  \frac{1024}3 T_3 &  \frac{272}9 T_1 - \frac{512}9T_3
\end{array}
\right)
 }}
where:
\eqa{
T_1(\delta) &=& \frac{23 \delta ^4}{16}-\frac{1}{2} \delta ^4 \log (\delta )-\frac{191 \delta ^3}{108}
+\frac{4}{9} \delta ^3 \log (\delta )+\frac{17\delta ^2}{18}-\frac{1}{3} \delta ^2 \log (\delta )+\frac{109 \delta }{18}\nn\\[1mm]
&&-\frac{5}{3} \delta  \log (\delta )+\frac{79}{18} \log(1-\delta )-\frac{5}{3} \log (1-\delta ) \log (\delta) \nn\\[1mm]
&&+\left[\delta ^4 -\frac{8}{9} \delta ^3 +\frac{2}{3} \delta ^2 +\frac{10}{3} \delta
+\frac{10}{3} \log (1-\delta)\right]\log \left(\frac{m_q}{m_b}\right) -\frac{5 \text{Li}_2(\delta )}{3}\ , \\[2mm]
T_2(\delta) &=& \frac{1181 \delta ^4}{2592}-\frac{17}{108} \delta ^4 \log (\delta )-\frac{395 \delta ^3}{648}
+\frac{4}{27} \delta ^3 \log (\delta)+\frac{7 \delta ^2}{18}-\frac{1}{9} \delta ^2 \log (\delta )+\frac{187 \delta }{108}\nn\\[1mm]
&&-\frac{1}{2} \delta  \log (\delta)+\frac{133}{108} \log (1-\delta )-\frac{1}{2} \log (1-\delta ) \log (\delta )\nn\\[1mm]
&&+\left[\frac{17 \delta ^4}{54}-\frac{8 \delta^3}{27}+\frac{2 \delta ^2}{9}
+\delta +\log (1-\delta )\right]\log \left(\frac{m_q}{m_b}\right)-\frac{\text{Li}_2(\delta )}{2}\ , \\[2mm]
T_3(\delta) &=& \frac{341 \delta ^4}{7776}-\frac{5}{324} \delta ^4 \log (\delta )-\frac{89 \delta ^3}{1944}
+\frac{1}{81} \delta ^3 \log (\delta)+\frac{\delta ^2}{72}-\frac{1}{108} \delta ^2 \log (\delta )+\frac{35 \delta }{162}\nn\\[1mm]
&&-\frac{1}{18} \delta  \log (\delta)+\frac{13}{81} \log (1-\delta )-\frac{1}{18} \log (1-\delta ) \log (\delta) \nn\\[1mm]
&&+\left[\frac{5 \delta ^4}{162}-\frac{2 \delta^3}{81}+\frac{\delta ^2}{54}+\frac{\delta }{9}
+\frac{1}{9} \log (1-\delta )\right]\log \left(\frac{m_q}{m_b}\right)-\frac{\text{Li}_2(\delta )}{18}\ .
}
The NLO matrix $G^{(1)}$ contains perturbative strong phases from on-shell contributions from light quarks,
as well as from charm quarks when the photon-energy cut is low enough. The matrix $G^{(1)}$ is the main result of this paper. It has the following structure:
\eq{
G^{(1)}(\mu,\delta) = G_1(\delta) L_q L_\mu +G_2(\delta) L_\mu + G_3(z_c,\delta) L_q
+ G_4(z_c,\delta)\ ,
}
where $L_\mu\equiv \log(\mu/m_b)$, $L_q\equiv \log(m_q/m_b)$ and $z_c\equiv m_c^2/m_b^2$. The
explicit form of $G^{(1)}_{ij}$ is too complicated to be written down here. However, it can be
constructed completely from the expressions in Sections~\ref{sec:Diags},~\ref{sec:renormalization} and
Appendix~\ref{sec:interresults}: Start from Eq.~(\ref{eq:master}), substitute the objects ${\cal D}_{(J)}$ from Eqs.~(\ref{eq:(i)}),~(\ref{eq:(ii)}),~(\ref{eq:(iii)}), then use Eqs.~(\ref{eq:P48})-(\ref{eq:P43})
and~(\ref{eq:P4C}) for the different matrix elements $\av{P_4}_{(J)}$, and use the expressions in the
appendix for the functions ${\cal F}_{(J)}$, ${\cal F}_{\text{coll}(J)}$ and $\widetilde {\cal F}_{(J)}$, noting
that $\hat {\cal F}_{(J)} \equiv {\cal F}_{(J)}+{\cal F}_{\text{coll}(J)}$. Finally, perform the replacement
$G^{(1)}\to G^{(1)} + G^{(1)\dagger}$ to account for the ``mirror" contributions. For convenience, we
provide the full matrices $G_{ij}^{(0)}$ and $G_{ij}^{(1)}$ in the file ``{\tt Gij.m}" attached to the arXiv submission
of the present manuscript.
The first is given by the $6\times 6$ matrix ``{\tt GijLO}" ($i,j=1u,2u,3,..,6$) and the second by the $10\times 10$ matrix
``{\tt GijNLO}" (with $i,j=1u,2u,3,..,6,1c,2c,7,8$).

\section{Numerical analysis}
\label{sec:numerics}

We briefly discuss here the numerical impact of the four-body contributions to the total $\bsg$ rate. We consider for convenience the following quantity:
\eq{
\widetilde{\Delta\Gamma} = \frac{\Delta \Gamma (\bar B\to X_s\gamma)_{E_\gamma>E_0}^{s\bar q q\gamma}}{\Gamma_0 \, \big|\C7^{(0)\text{eff}}\big|^2},
}
given by Eq.~(\ref{eq:DeltaGamma}) and normalized to the leading contribution to the decay rate.
The Wilson coefficients are given by:
\eq{
\C{i}^\text{eff}(\mu) = \C{i}^{(0)\text{eff}}(\mu) + \frac{\alpha_s(\mu)}{4\pi} \C{i}^{(1)\text{eff}}(\mu) + \ord{\alpha_s^2},
}
which are computed following Ref.~\cite{9910220}. For the reference matching and renormalization scales
$\mu_0 = 160$ GeV, $\mu = \mu_b = 2.5$ GeV, we have\footnote{
The NLO Wilson Coefficients $\C{7,8}^\text{eff}$ are not needed for our NLO results as $P_{7,8}$ do not contribute at LO.
}:
\eqa{
\C{i}^{(0)\text{eff}} &=&
(0.828 \,\lambda_q,-1.063\,\lambda_q,-0.013,-0.125,0.0012,0.0027,-0.372,-0.172)\ ,\quad\\
\C{i}^{(1)\text{eff}} &=&
(-15.32\,\lambda_q,2.10\,\lambda_q,	0.097,-0.447,-0.021,-0.013,\text{nn},\text{nn})\ ,
}
for $i= 1q,2q,3,..,8$. However, the $\mu$-dependence of the Wilson coefficients is important and we will analyze it here. In addition,
$\lambda_q \equiv V_{qs}^{*}V_{qb}/V_{ts}^{*}V_{tb}$ denote the appropriate CKM factors,
given by \cite{pdg}:
\eq{
\lambda_u =  -0.0059 + 0.018 i \ ,\quad
\lambda_c = -0.97 \ .
}
The quantity $\widetilde{\Delta\Gamma}$ can be expanded in $\alpha_s$:
\eqa{\label{tildeDeltaGamma}
\widetilde{\Delta\Gamma} &=& \widetilde{\Delta\Gamma}_{\rm LO} + \widetilde{\Delta\Gamma}_{\rm NLO}
= \sum_{
\begin{minipage}{16mm}
\scriptsize\flushright
$i,j=1u,2u$\\[-1mm]$3,..,6$
\end{minipage}
} \frac{\C{i}^{(0){\vtwo{\text{eff}}}*}\C{j}^{(0){\vtwo{\text{eff}}}}}{\big|\C7^{(0)\text{eff}}\big|^2} \,G_{ij}^{(0)}
\label{eq:DeltaGexp}\\[2mm]
&&+ \frac{\alpha_s(\mu)}{4\pi} \left[ \sum_{
\begin{minipage}{16mm}
\scriptsize\flushright
$i,j=1u,2u$\\[-1mm]$3,..,6$
\end{minipage}
}
\frac{\C{i}^{(1){\vtwo{\text{eff}}}*}\C{j}^{(0){\vtwo{\text{eff}}}}+\C{i}^{(0){\vtwo{\text{eff}}}*}\C{j}^{(1){\vtwo{\text{eff}}}}}{\big|\C7^{(0)\text{eff}}\big|^2} \,G_{ij}^{(0)}
+ \sum_{i,j=all} \frac{\C{i}^{(0){\vtwo{\text{eff}}}*}\C{j}^{(0){\vtwo{\text{eff}}}}}{\big|\C7^{(0)\text{eff}}\big|^2} \,G_{ij}^{(1)}
\right] \ .\nn
}
We begin with a discussion of the $\mu$-dependence of our results. To leading order, the $\mu$-dependence
is given purely by the LL (leading-log) running in the effective theory. Note that to this order, only $\C{1,2}^u$
and $\C{3,4,5,6}$ contribute. At NLO, three new contributions arise: (i) the contribution from NLO Wilson 
coefficients, (ii) NLO matrix elements and (iii) contributions from $\C{1,2}^c,\C{7,8}$, absent at LO. The
$\mu$-dependence should cancel up to a residual scale-dependence from higher orders, and up to
the neglected contributions shown in Fig.~\ref{fig:examples}.f (note that the $Z$ factors in
Eq.~(\ref{eq:Zfactors}) are not the full renormalization constants).

\begin{figure}
\includegraphics[width=7.2cm,height=5cm]{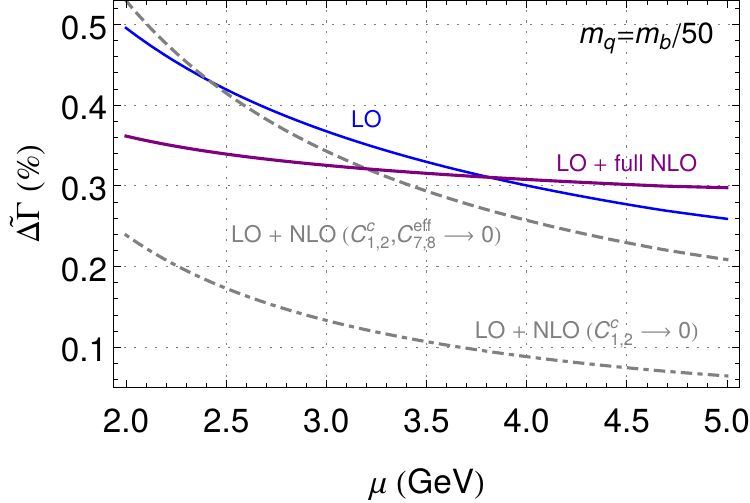}\hspace{5mm}
\includegraphics[width=7.3cm,height=5cm]{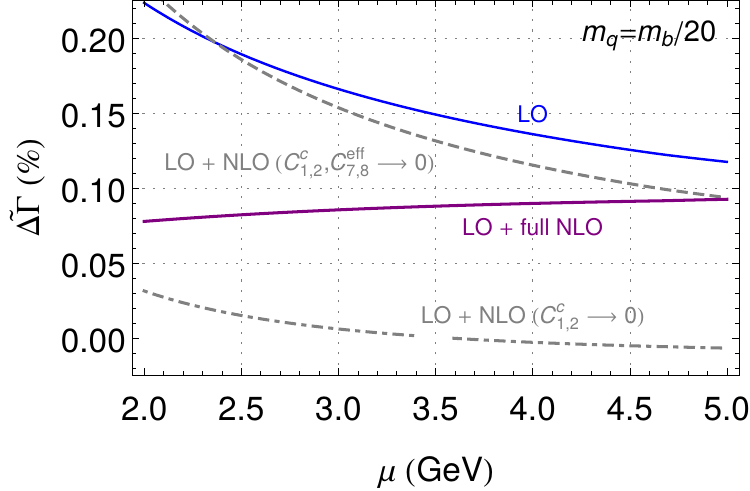}
\caption{
Renormalization-scale dependence of $\widetilde{\Delta\Gamma}$ in \emph{percent units}.
Here we have taken $\mu_0=160$ GeV, $z_c = 0.07$, $\delta=0.316$ and $L_q=-\log{50}$ ($m_q\sim 100$ MeV) [Left panel], or $L_q=-\log{20}$ ($m_q\sim 250$ MeV) [Right panel].
}
\label{fig:plotmu}
\end{figure}

In Fig.~\ref{fig:plotmu} we show the $\mu$-dependence of the LO result, and LO+NLO
excluding $\C{1,2}^c,\C{7,8}$, LO+NLO excluding $\C{1,2}^c$ and LO+ full NLO.
We also gauge the impact of collinear logarithms, showing the result for two
different choices of $L_q$, corresponding to $m_q=m_b/50$ ($m_q\sim 100$ MeV) and
$m_q=m_b/20$ ($m_q\sim 250$ MeV). Collinear logarithms are, as expected, numerically important. 

Contributions from $P_{1,2}^c$ and $P_{7,8}$ arise only at NLO and therefore introduce at this order a novel
$\mu$-dependence. Although, as we will see, certain cancellations make the NLO contribution small, there is
a considerable reduction in the renormalization-scale dependence of the full LO+NLO result as compared to the LO contribution alone. This is due to the fact that the main $\mu$-dependence of the leading order contribution
arises from the mixing of $P_{1,2}^c$ onto penguin operators, which is compensated at NLO by the matrix
elements of $P_{1,2}^c$. This can be seen in Fig.~\ref{fig:plotmu}: the reduction in the $\mu$-dependence
is achieved only after including $\C{1,2}^c$ contributions.

In the left plot of Fig.~\ref{fig:plotmu} one can see that for the value $\mu\simeq 4$ GeV strong cancellations make the
NLO contribution very small. More concretely, for the inputs $\mu_0=160$ GeV, $\mu=4$ GeV, $z_c=0.07$, $\delta=0.316$
and $m_q=m_b/50$, we have:
\eqa{
\widetilde{\Delta\Gamma} (\%)
&=& (0.300)_\text{\tiny LO} + (0.044)_\text{\tiny NLO WCs} - (0.087)_\text{\tiny NLO penguins}
- (0.169)_{\scriptstyle \C{7,8}^\text{eff}} + (0.219)_{\scriptstyle \C{1,2}^c} \nn\\[2mm]
&=& (0.300)_\text{\tiny LO} + (0.044)_\text{\tiny NLO WCs}  - (0.036)_\text{\tiny NLO MEs}\nn\\[2mm]
&=& (0.300)_\text{\tiny LO} - (0.007)_\text{\tiny NLO}
}
where the term labeled `NLO WCs' corresponds to the second term in Eq.~(\ref{tildeDeltaGamma}).
This cancellation is very efficient for $\mu\simeq 3.8$ GeV, but depends strongly on $m_q$ and $z_c$.
However, it is a general feature of our results that the contribution from $\C{1,2}^c$ is of the same
order as the rest of the NLO contribution, but with opposite sign, leading always to some level of cancellation.
Note also that the (NLO) $\C{1,2}^c$ contribution is also as large as the LO result.

\begin{figure}
\includegraphics[width=7.2cm,height=5cm]{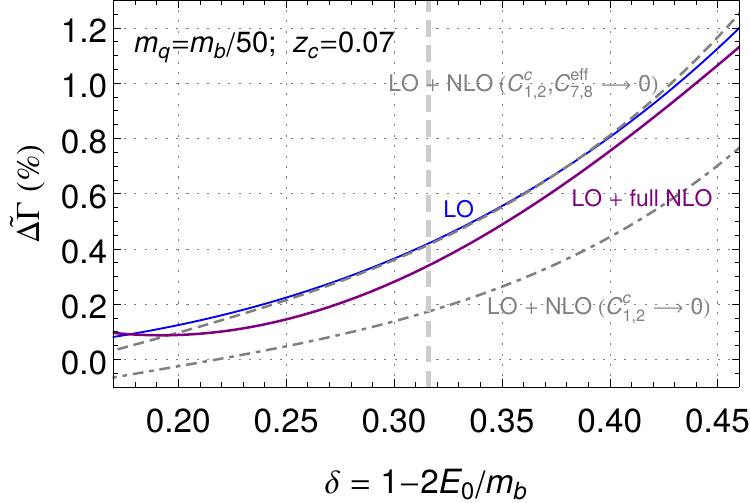}\hspace{5mm}
\includegraphics[width=7.3cm,height=5cm]{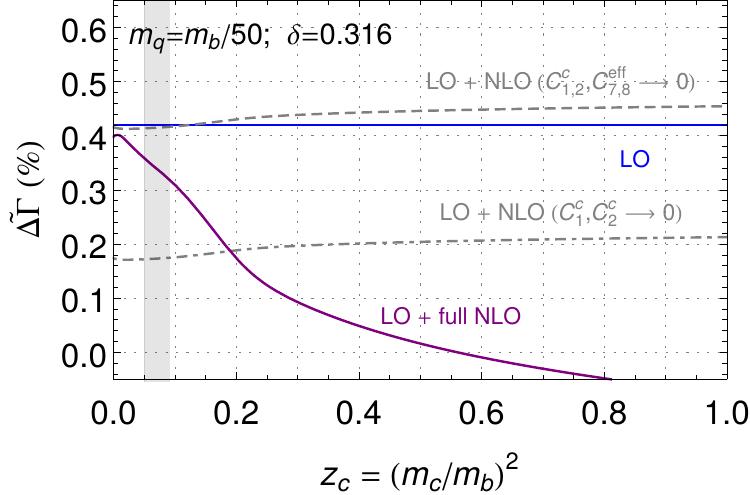}
\caption{
$\widetilde{\Delta\Gamma}$ in \emph{percent units}.
Left: Dependence on the photon energy cut $E_0$.
Right: Dependence on the charm mass. We have fixed $\mu_0=160$ GeV, $\mu=2.5$ GeV and $L_q=-\log{50}$ ($m_q\sim 100$ MeV). The vertical dashed line in the left panel shows the benchmark point $\delta=0.316$, while the vertical
band in the right panel corresponds to the physical value $z_c=0.07\pm 0.02$.
}
\label{fig:plotdeltazc}
\end{figure}

In the following we fix the renormalization scale to $\mu=2.5$ GeV and study the dependence on the charm mass and the
photon-energy cut. This is shown in Fig.~\ref{fig:plotdeltazc}. In general the full LO+NLO result increases with $\delta$
and decreases with $m_c$, always remaining below the $1\%$ level for $\delta\lesssim 0.45$. We note that these results
are only valid for $\delta$ not far from $0.316$ as some of the functions are expanded up to second order in
$(\delta-0.316)$.
 
Finally, we provide some results for two different values of $E_0$ of interest: $E_0=1.6$~GeV, corresponding to $\delta=0.316$,
and $E_0=1.9$~GeV, corresponding to $\delta=0.188$. For the input parameters and their uncertainties we take:
$\mu_0=160^{+90}_{-80}$, $\mu=2.5^{+2.5}_{-0.5}$ and $z_c=0.07\pm 0.02$, which captures the different values for $m_c$
within different schemes.\\[-3mm]

\noindent For $m_q=m_b/50\sim 100$ MeV, we find:
\eqa{
\widetilde{\Delta\Gamma}_{E_0=1.6\text{ GeV}}\ [\%] &=&
(0.419)_\text{\tiny LO} - (0.080)_\text{\tiny NLO}
\pm (0.028)_{\mu_0}  \pm (0.032)_{\mu} \pm (0.019)_{z_c} \nn\\[1mm]
&=& 0.34 \pm 0.05 \\[2mm]
\widetilde{\Delta\Gamma}_{E_0=1.9\text{ GeV}}\ [\%] &=&
(0.105)_\text{\tiny LO} - (0.077)_\text{\tiny NLO}
\pm (0.012)_{\mu_0}  \pm (0.009)_{\mu} \pm (0.003)_{z_c} \nn\\[1mm]
&=& 0.03 \pm 0.02
\label{E0=1.9,Lq1/50}
}
For $m_q=m_b/20\sim 250$ MeV:
\eqa{
\widetilde{\Delta\Gamma}_{E_0=1.6\text{ GeV}}\ [\%] &=&
(0.189)_\text{\tiny LO} - (0.107)_\text{\tiny NLO}
\pm (0.019)_{\mu_0}  \pm (0.007)_{\mu} \pm (0.007)_{z_c} \nn\\[1mm]
&=& 0.08 \pm 0.02\\[2mm]
\widetilde{\Delta\Gamma}_{E_0=1.9\text{ GeV}}\ [\%] &=&
(0.037)_\text{\tiny LO} - (0.081)_\text{\tiny NLO}
\pm (0.009)_{\mu_0}  \pm (0.020)_{\mu} \pm (0.001)_{z_c} \nn\\[1mm]
&=& -0.04 \pm 0.02
\label{E0=1.9,Lq1/20}
}

For the value $\delta=0.188$ ($E_0=1.9$ GeV) we have used the exact results (not the expanded ones), as for
this value of $\delta$ the quadratic expansion is not expected to be accurate enough.

\skipp{
For example, the expanded expressions
would give $-(0.018)_\text{\tiny NLO}$ instead of $-(0.076)_\text{\tiny NLO}$ in Eq.~(\ref{E0=1.9,Lq1/50}), and
$-(0.041)_\text{\tiny NLO}$ instead of $-(0.080)_\text{\tiny NLO}$ in Eq.~(\ref{E0=1.9,Lq1/20}).
}


\section{Conclusions}
\label{sec:conclusions}

The inclusive radiative decay $\bsg$ has beyond any doubt reached the era of precision physics, with the total uncertainties
on both the experimental and theoretical side being at the $\pm 7\%$ level. The foreseen improvement in precision on the experimental side --~the envisaged uncertainty with $50/$ab at Belle II is of ${\cal O}(6\%)$~\cite{1011.0352}, although this might even be a conservative estimate~-- 
justifies every effort to reduce the theoretical error to at least the same level.

The present article aims at addressing a particular higher-order perturbative contribution, namely the four-body contributions
$b\to s q\bar q \gamma$ to $\Gamma(\bsg)$ at NLO. The smallness of the Wilson coefficients of penguin operators and
CKM-suppression of current-current operators suggests that this contribution should be small. However, only an explicit
calculation can turn this estimate into a firm statement. The calculation arises from tree and one-loop amplitudes,
but it involves the four-body phase-space integration in dimensional regularization, which makes the calculation non-trivial
owing to the appearance of higher transcendental functions. Moreover, the cancellation of poles in the dimensional regularization
parameter $\epsilon$ is only achieved after proper UV and IR renormalization. The latter gives rise to logarithms
$\ln(m_q/m_b)$ when turning the dimensional into a mass regulator. These logarithms stem from the phase space region of
energetic collinear photon radiation off light quarks in the final state. They are computed with the splitting
function technique and treated in the same way as in~\cite{0512066,1209.0965}.

We find indeed that the contribution of our four-body NLO correction to the total rate is below the $1\%$ level, as expected.
This statement even holds true once we vary the input parameters such as the charm mass $m_c$, the photon energy cut
(parameterized by $\delta$), the masses $m_q$ of the light quarks, or the renormalization and matching scales, as can be seen
by the numbers and the plots in Section~\ref{sec:numerics}. We also confirm the LO results presented previously in
Ref.~\cite{1209.0965}.

Yet the NLO calculation of $\bsg$ is still not complete. There are certain yet unpublished 
three-particle cuts contributing to $\Gamma(b\to sg\gamma)$, mainly interferences of $P_{1,2}^u$ with $P_{1,2}^c$, which are also of the $(A,B)$-interference type.
These contributions can be extracted from the results of Ref.~\cite{9512252}. The only missing pieces are given by the diagrams in Fig.~\ref{fig:examples}.f. These are NLO interferences
of the type $(B,B)$ and are expected to be negligible with respect to the $(A,B)$ ones that we have calculated in a complete manner.
While these contributions can be calculated with the techniques described in this paper, they are left for future work.

\section*{Acknowledgements}
We thank Mikolaj Misiak for valuable correspondence and comments on the manuscript,
and Dirk Seidel for useful discussions about penguins.
This work has been funded by the Deutsche Forschungsgemeinschaft (DFG) within research unit FOR 1873 (QFET). M.P. acknowledges support by the National Science Centre (Poland)
research project, decision DEC-2011/01/B/ST2/00438.


\appendix

\section{Intermediate Results}
\label{sec:interresults}

\subsection{$(P_7,P_i)$ interference}

The functions ${\cal F}^{I,7}_{(J)}(\delta)$ are given by:

\eqa{
{\cal F}^{s,7}_{(i)}(\delta)&=& -6\, {\cal F}^{\times,7}_{(i)}(\delta)\ ;\\[4mm]
{\cal F}^{\times,7}_{(i)}(\delta)&=&
\frac{-18\delta + 33 \delta^2 - 2 \delta^3 - 13 \delta^4 + 6 \delta^3 (2 + \delta)\log(\delta)}{243 (1-\delta)}\ ;\\[4mm]
{\cal F}^{\times,7}_{(ii)}(\delta)&=&
\frac{ 12\delta - 3 \delta^2 - 8 \delta^3 - \delta^4 + 6 \delta^2 (2 + \delta)\log(\delta)}{54 (1-\delta)}\ ;
}

\subsection{$(P_8,P_i)$ interference}

Up to subleading terms in $\ep$, we have always
\eq{
{\cal F}^{s,8}(\delta)= -6 \,(1+\ep)\, {\cal F}^{\times,8}(\delta) \ .
}
The functions ${\cal F}^{\times,8}(\delta)$ are given by:
\allowdisplaybreaks{
\eqa{
{\cal F}^{\times,8}_{(i)}(\delta)&=&
A_8(\delta) \bigg[\frac1{\ep} + 6 L_\mu \bigg] + B_8(\delta)\ ; \\[4mm]
{\cal F}^{\times,8}_{(ii)}(\delta) &=& B_8'(\delta)\ ;
\qquad\quad {\cal F}^{\times,8}_{\text{coll}(ii)}(\delta)= 0 \ ;\\[4mm]
{\cal F}^{\times,8}_{(iii)}(\delta)&=&
A_8''(\delta) \bigg[\frac1{\ep} + 6 L_\mu \bigg] + B_8''(\delta)\ ;\\[4mm]
{\cal F}^{\times,8}_{\text{coll}(i)}(\delta)&=&
-A_8(\delta) \left[\frac1{\ep} +6 L_\mu -2 L_q  \right] + D_8(\delta)\ ;\\[4mm]
{\cal F}^{\times,8}_{\text{coll}(iii)}(\delta)&=&
-A_8''(\delta) \left[\frac1{\ep} +6 L_\mu -2 L_q  \right] + D_8''(\delta)\ ;
}}\\
where $L_\mu = \log(\mu/m_b)$ and $L_q = \log(m_q/m_b)$, and:
\eqa{
A_8(\delta) &=& \frac{4 \delta ^3}{81}-\frac{\delta ^2}{27}+\frac{4 \delta }{27}+\frac{4}{27} \log (1-\delta)\ ;\\[4mm]
A_8''(\delta) &=& -\frac{4 \delta^3}{81}+\frac{10\delta^2}{27}+\frac{2\delta}{27}
-\left(\frac{2\delta^2}{9}-\frac{4\delta}{9}-\frac{2}{27} \right) \log (1-\delta)\ ;\\[4mm]
B_8(\delta) &=& \frac{62 \delta ^3}{243}-\frac{17 \delta ^2}{162}+\frac{116 \delta}{81}
-\left(\frac{8\delta ^3}{81}-\frac{2 \delta ^2}{27}+\frac{8 \delta}{27}\right) \log\delta
-\frac8{27}\log\delta\log (1-\delta)\nn\\[2mm]
&-&
\log (1-\delta ) \left(\frac{8\delta ^3}{81}-\frac{2 \delta ^2}{27}+\frac{8 \delta}{27} -\frac{92}{81} \right) 
-\frac4{27}\log^2 (1-\delta) - \frac{8 \text{Li}_2(\delta )}{27}\ ;\\[4mm]
B_8'(\delta)&=&
\frac{4\delta }{27}-\frac{\delta ^2}{9}+\frac{2\delta ^3}{81}+\frac{4\log (1-\delta)}{27}\ ;\\[4mm]
B_8''(\delta)&=&
-\frac{8 \delta ^3}{27}+\frac{199 \delta ^2}{81}+\frac{119 \delta }{81}
+\bigg( \frac{8 \delta ^3}{81}-\frac{20 \delta ^2}{27}-\frac{4\delta}{27} \bigg)\log\delta \nn\\[2mm]
&+&
\bigg( \frac{8 \delta ^3}{81}-\frac{47 \delta ^2}{27}+\frac{50 \delta }{27}+\frac{107}{81}
+\frac{4\delta^2\log\delta}9 - \frac{8 \delta \log\delta}9 -\frac{4\log\delta}{27} \bigg)\log(1-\delta)\nn\\[2mm]
&+&
\bigg( \frac{5 \delta ^2}{9}-\frac{10 \delta }{9}+\frac{7}{27} \bigg)\log^2(1-\delta)
+\bigg(\frac{4\delta^2}9 - \frac{8 \delta}9 - \frac{4}{27}\bigg) \text{Li}_2(\delta)\ ;\\[4mm]
D_8(\delta) &=&
-\frac{37 \delta ^3}{243}+\frac{\delta ^2}{54} -\frac{8 \delta }{9}
+\log (1-\delta) \left(\frac{8 \delta ^3}{81}-\frac{2 \delta ^2}{27}+\frac{8 \delta}{27}+\frac{4 \log\delta}{27}
-\frac{20}{27}\right)\nn\\[2mm]
&+&\left(\frac{4 \delta ^3}{81}-\frac{\delta ^2}{27}+\frac{4 \delta }{27}\right) \log (\delta)
+\frac{4}{27} \log ^2(1-\delta)+\frac{4 \text{Li}_2(\delta )}{27}\ ;\\[4mm]
D_8''(\delta) &=&
+\frac{37 \delta ^3}{243}-\frac{35 \delta ^2}{27}-\frac{10 \delta }{9}
- \left(\frac{8\delta ^3}{81}-\frac{32 \delta ^2}{27}+\frac{20\delta }{27}+\frac{28}{27}\right)\log (1-\delta)\nn\\[2mm]
&-&
\left(\frac{4 \delta ^3}{81}-\frac{10 \delta^2}{27}-\frac{2 \delta }{27}\right) \log\delta
-\left(\frac{2\delta^2}{9}-\frac{4 \delta}{9}-\frac{2}{27}\right) \log\delta\log (1-\delta)
\nn \\[2mm]
&-&
\left(\frac{4 \delta ^2}{9}-\frac{8 \delta }{9}+\frac{4}{27}\right) \log ^2(1-\delta)
-\left( \frac{2 \delta ^2}{9}-\frac{4 \delta}{9}-\frac{2}{27} \right) \text{Li}_2(\delta) \ ;
}

\subsection{$(P_i,P_j)$ interference}

For ${\cal F}^{I,1}_{(J)}(\delta)$ we give analytical results for $m_c$-independent functions, but the $m_c$-dependence is given as interpolated functions.
Up to subleading terms in $\ep$, we have always
\eq{
{\cal F}^{s}(z_q,\delta)=
-6 \,(1+\ep+\ep^2)\, {\cal F}^{\times}(z_q,\delta) \ .
}
The functions ${\cal F}^{\times}(z_q,\delta)$ are given by:

\eqa{
{\cal F}^{\times,1}_{(i)}(z_q,\delta)&=&
A(\delta) \bigg[\frac1{\ep^2} + \frac1{\ep} 8 L_\mu +32 L_\mu^2 \bigg]
+B(z_q,\delta) \bigg[\frac1{\ep} + 8 L_\mu \bigg] + C(z_q,\delta)\ ;\qquad \\[4mm]
{\cal F}^{\times,1}_{(ii)}(z_q,\delta)&=&
B'(\delta) \bigg[\frac1{\ep} + 8 L_\mu \bigg] + C'(z_q,\delta)\ ;\qquad 
{\cal F}^{\times,1}_{\text{coll}(ii)}(z_q,\delta) = 0\ ; \\[2mm]
{\cal F}^{\times,1}_{(iii)}(z_q,\delta)&=&
A''(\delta) \bigg[\frac1{\ep^2} + \frac1{\ep} 8 L_\mu +32 L_\mu^2 \bigg]
+B''(z_q,\delta) \bigg[\frac1{\ep} + 8 L_\mu \bigg] + C''(z_q,\delta)\ ;\qquad \\[4mm]
{\cal F}^{\times,1}_{\text{coll}(i)}(z_q,\delta)&=&
-A(\delta)\, \bigg[\frac1{\ep^2} + \frac1{\ep} (8 L_\mu - 2 L_q) + 30 L_\mu^2 - 12 L_\mu L_q + 2 L_\mu - 2 L_q \bigg]\\[2mm]
&&
\hspace{-15mm}
-[B(z_q,\delta)+F(\delta)+H(\delta)]\, \bigg[\frac1{\ep} + 8 L_\mu - 2 L_q \bigg] + f(\delta)\, [L_\mu-L_q] + E(z_q,\delta) \ ;\qquad \nn\\[4mm]
{\cal F}^{\times,1}_{\text{coll}(iii)}(z_q,\delta)&=&
-A''(\delta)\, \bigg[\frac1{\ep^2} + \frac1{\ep} (8 L_\mu - 2 L_q) + 30 L_\mu^2 - 12 L_\mu L_q + 2 L_\mu - 2 L_q \bigg] \\[2mm]
&&
\hspace{-15mm}
-[B''(z_q,\delta)\!+F''(\delta)\!+H''(\delta)]\, \bigg[\frac1{\ep} + 8 L_\mu - 2 L_q \bigg] + f''(\delta)\, [L_\mu-L_q] + E''(z_q,\delta) \ ;\qquad\nn 
}
Counterterms are given by:

\eqa{
{\cal F}^{\times,C}_{(i)}(\delta)&=&
-A(\delta) \bigg[\frac1{\ep^2} + \frac1{\ep} 6 L_\mu + 18 L_\mu^2 \bigg]
+F(\delta) \bigg[\frac1{\ep} + 6 L_\mu \bigg] + G\ ;\qquad \\[3mm]
{\cal F}^{\times,C}_{(ii)}(\delta)&=&
-B'(\delta) \bigg[\frac1{\ep} + 6 L_\mu \bigg] + G'(\delta)\ ;\qquad
{\cal F}^{\times,C}_{\text{coll}(ii)}(z_q,\delta) = 0\ ;\\[3mm]
{\cal F}^{\times,C}_{(iii)}(\delta)&=&
-A''(\delta) \bigg[\frac1{\ep^2} + \frac1{\ep} 6 L_\mu + 18 L_\mu^2 \bigg]
+F''(\delta) \bigg[\frac1{\ep} + 6 L_\mu \bigg] + G''(\delta)\ ;\qquad \\[4mm]
{\cal F}^{\times,C}_{\text{coll}(i)}(\delta)&=&
A(\delta)\, \bigg[\frac1{\ep^2} + \frac1{\ep} (6 L_\mu - 2 L_q) + 16 L_\mu^2 - 8 L_\mu L_q + 2 L_\mu - 2 L_q \bigg]\nn\\[2mm]
&&
+H(\delta)\, \bigg[\frac1{\ep} + 6 L_\mu - 2 L_q \bigg] - f(\delta)\, [L_\mu-L_q] + I(\delta) \ ;\qquad \\[3mm]
{\cal F}^{\times,C}_{\text{coll}(iii)}(\delta)&=&
A''(\delta)\, \bigg[\frac1{\ep^2} + \frac1{\ep} (6 L_\mu - 2 L_q) + 16 L_\mu^2 - 8 L_\mu L_q + 2 L_\mu - 2 L_q \bigg]\nn\\[2mm]
&&
+H''(\delta)\, \bigg[\frac1{\ep} + 6 L_\mu - 2 L_q \bigg] - f''(\delta)\, [L_\mu-L_q] + I''(\delta) \ ;\qquad
}
where again $L_\mu = \log(\mu/m_b)$ and $L_q = \log(m_q/m_b)$. From these expressions one can check the pattern of
cancellation of UV and collinear divergences.
The $z_q$-independent functions are given by (with the notation $\bar\delta \equiv 1-\delta$, $L_{\bar\delta}\equiv\log(1-\delta)$,
$L_\delta\equiv\log\delta$),

{\small
\eqa{
A(\delta)&=&
-\frac{\delta ^4}{1458}-\frac{\delta}{243}-\frac{1}{243} L_{\bar{\delta }}\ ;\\[5mm]
A''(\delta)&=&
-\frac{2 \delta ^4}{729}+\frac{2 \delta ^3}{729}-\frac{\delta ^2}{486}-\frac{2 \delta }{243} -\frac{2}{243} L_{\bar{\delta }}\ ;\\[5mm]
B'(\delta)&=&
-\frac{2 \delta ^3}{2187}+\frac{2 \delta ^2}{729}-\frac{2 \delta }{729}-\frac{2}{729} L_{\bar{\delta }}\ ;\\[5mm]
f(\delta)&=&
\left(\frac{4 \delta ^2}{243}-\frac{4 \delta }{729}-\frac{1}{243}\right) L_{\bar{\delta }}+\frac{2}{243}L_{\bar{\delta }}^2+\frac{11
   \delta ^4}{8748}+\frac{5 \delta ^3}{729}-\frac{23 \delta ^2}{1458}-\frac{\delta }{243}
\ ;\\[5mm]
f''(\delta)&=&
\left(\frac{10 \delta ^2}{243}-\frac{16 \delta }{729}-\frac{2}{729}\right) L_{\bar{\delta }}+\frac{4}{243}L_{\bar{\delta }}^2+\frac{11
   \delta ^4}{2187}+\frac{34 \delta ^3}{2187}-\frac{29 \delta ^2}{729}-\frac{2 \delta }{729}
\ ;\\[5mm]
F(\delta)&=&
-\left(\frac{2 L_{\bar{\delta }}}{243}+\frac{\delta ^4}{729}+\frac{2 \delta }{243}\right)L_{\delta } -\left(\frac{\delta ^4}{729}+\frac{2
   \delta }{243}-\frac{97}{2916}\right) L_{\bar{\delta }}-\frac{L_{\bar{\delta }}^2}{243}+\frac{53 \delta ^4}{17496}+\frac{4 \delta
   ^3}{2187}\nn\\
&&+\frac{7 \delta ^2}{5832}+\frac{121 \delta }{2916}-\frac{2 \text{Li}_2(\delta )}{243}
\ ;\\[6mm]
F''(\delta)&=&
-\left(\frac{4 L_{\bar{\delta }}}{243}+\frac{4 \delta ^4}{729}-\frac{4 \delta ^3}{729}+\frac{\delta ^2}{243}+\frac{4 \delta}{243}\right)L_{\delta}
   -\left(\frac{4 \delta ^4}{729}-\frac{4 \delta ^3}{729}+\frac{\delta ^2}{243}+\frac{4 \delta }{243}-\frac{47}{729}\right)
   L_{\bar{\delta }}\nn\\
&&-\frac{2 L_{\bar{\delta }}^2}{243}+\frac{71 \delta ^4}{4374}-\frac{38 \delta ^3}{2187}+\frac{49 \delta
   ^2}{2916}+\frac{59 \delta }{729}-\frac{4 \text{Li}_2(\delta )}{243}
\ ;\\[6mm]
G(\delta)&=&
\left(\frac{\delta ^4}{729}+\frac{2 \delta }{243}-\frac{97}{2916}\right) L_{\bar{\delta }}^2
+\left(\frac{2L_{\bar{\delta }}}{243}+\frac{\delta ^4}{729}+\frac{2 \delta }{243}\right)L_{\delta}^2
+\frac{2 L_{\bar{\delta }}^3}{729}-\left(\frac{53 \delta
   ^4}{8748}+\frac{8 \delta ^3}{2187}\right.\nn\\
&&\left.+\frac{7 \delta ^2}{2916}+\frac{121 \delta }{1458}-\frac{4 \text{Li}_2(\delta )}{243}+\frac{7 \pi
   ^2}{972}-\frac{6901}{34992}\right) L_{\bar{\delta }} + \left(\frac{2 \delta ^4}{729}+\frac{4 \delta
   }{243}-\frac{97}{1458}\right) L_{\delta} L_{\bar{\delta}}\nn\\
&&+\left(\frac{4 L_{\bar{\delta }}^2}{243}-\frac{53 \delta ^4}{8748}-\frac{8 \delta^3}{2187}-\frac{7 \delta ^2}{2916}-\frac{121 \delta }{1458}
+\frac{4 \text{Li}_2(\delta)}{243}\right)L_{\delta}-\frac{13\pi^2\delta^4}{17496}+\frac{2233 \delta ^4}{209952}+\frac{\delta ^3}{81} \nn\\
&&+\frac{733 \delta ^2}{69984}-\frac{13 \pi ^2 \delta }{2916}+\frac{9805
   \delta }{34992}-\frac{97 \text{Li}_2(\delta )}{1458}+\frac{4 \text{Li}_3(\bar\delta )}{243}-\frac{4 \text{Li}_3(\delta )}{243}-\frac{4
   \zeta(3)}{243}
\ ;\\[6mm]
G'(\delta)&=&
 -\left(\frac{4 L_{\bar{\delta }}}{729}+\frac{4 \delta ^3}{2187}-\frac{4 \delta ^2}{729}+\frac{4 \delta
   }{729}\right)L_{\delta }-\left(\frac{4 \delta ^3}{2187}-\frac{4 \delta ^2}{729}+\frac{4 \delta }{729}-\frac{59}{2187}\right) L_{\bar{\delta
   }}-\frac{2 L_{\bar{\delta }}^2}{729}\nn\\
&&+\frac{5 \delta ^3}{729}-\frac{50 \delta ^2}{2187}+\frac{71 \delta }{2187}-\frac{4\text{Li}_2(\delta )}{729} \ ;
}
\eqa{
G''(\delta)&=&
\left(\frac{4 \delta ^4}{729}-\frac{4 \delta ^3}{729}+\frac{\delta ^2}{243}+\frac{4 \delta }{243}-\frac{47}{729}\right) L_{\bar{\delta}}^2
+\left(\frac{4 L_{\bar{\delta }}}{243}+\frac{4 \delta ^4}{729}-\frac{4 \delta ^3}{729}+\frac{\delta^2}{243}+\frac{4 \delta}{243}\right)L_{\delta}^2 \nn\\
&&+\frac{4}{729}L_{\bar{\delta }}^3-\left(\frac{71 \delta ^4}{2187}-\frac{76 \delta
   ^3}{2187}+\frac{49 \delta ^2}{1458}+\frac{118 \delta }{729}-\frac{8 \text{Li}_2(\delta )}{243}+\frac{7 \pi
   ^2}{486}-\frac{1645}{4374}\right) L_{\bar{\delta }} \nn\\
&&-L_{\delta } \left[\left(-\frac{8 \delta ^4}{729}+\frac{8 \delta ^3}{729}-\frac{2
   \delta ^2}{243}-\frac{8 \delta }{243}+\frac{94}{729}\right) L_{\bar{\delta }}-\frac{8 L_{\bar{\delta }}^2}{243}+\frac{71 \delta
   ^4}{2187}-\frac{76 \delta ^3}{2187}+\frac{49 \delta ^2}{1458} \right.\nn\\
&&\left.+\frac{118 \delta }{729}-\frac{8 \text{Li}_2(\delta
   )}{243}\right]-\frac{13 \pi ^2 \delta ^4}{4374}+\frac{1877 \delta ^4}{26244}+\frac{13 \pi ^2 \delta ^3}{4374}-\frac{527 \delta
   ^3}{6561}-\frac{13 \pi ^2 \delta ^2}{5832}+\frac{1493 \delta ^2}{17496}\nn\\
&&-\frac{13 \pi ^2 \delta }{1458}+\frac{2353 \delta
   }{4374}-\frac{94 \text{Li}_2(\delta )}{729}+\frac{8 \text{Li}_3(\bar\delta )}{243}-\frac{8 \text{Li}_3(\delta )}{243}-\frac{8 \zeta
   (3)}{243}
\ ;\\[8mm]
H(\delta)&=&
\left(\frac{L_{\bar{\delta }}}{243}+\frac{\delta ^4}{1458}+\frac{\delta }{243}\right)L_{\delta}+\left(\frac{\delta ^4}{729}+\frac{2
   \delta }{243}-\frac{7}{324}\right) L_{\bar{\delta }}+\frac{L_{\bar{\delta }}^2}{243}-\frac{5 \delta ^4}{3888}-\frac{19 \delta
   ^3}{8748} \nn\\
&&-\frac{7 \delta ^2}{5832}-\frac{25 \delta }{972}+\frac{\text{Li}_2(\delta )}{243}
\ ;\\[8mm]
H''(\delta)&=&
\left(\frac{2 L_{\bar{\delta }}}{243}+\frac{2 \delta ^4}{729}-\frac{2 \delta ^3}{729}+\frac{\delta ^2}{486}+\frac{2 \delta
   }{243}\right)L_{\delta}+\left(\frac{4 \delta ^4}{729}-\frac{4 \delta ^3}{729}+\frac{\delta ^2}{243}+\frac{4 \delta }{243}-\frac{7}{162}\right)
   L_{\bar{\delta }}\nn\\
&&+\frac{2 L_{\bar{\delta }}^2}{243}-\frac{2 \delta ^4}{243}+\frac{29 \delta ^3}{4374}-\frac{8 \delta
   ^2}{729}-\frac{25 \delta }{486}+\frac{2 \text{Li}_2(\delta )}{243}
\ ;\\[8mm]
I(\delta)&=&
-\left(\frac{L_{\bar{\delta }}}{486}+\frac{\delta ^4}{2916}+\frac{\delta }{486}\right)L_{\delta}^2-\left(\frac{\delta ^4}{729}-\frac{2
   \delta ^2}{243}+\frac{8 \delta }{729}-\frac{23}{972}\right) L_{\bar{\delta }}^2+\left(\frac{5 \delta ^4}{1944}+\frac{19 \delta
   ^3}{4374} \right.\nn\\
&&\left.+\frac{67 \delta ^2}{2916}+\frac{61 \delta }{4374}+\frac{2 \text{Li}_2(\bar\delta )}{243}+\frac{\pi
   ^2}{486}-\frac{809}{11664}\right) L_{\bar{\delta }}-L_{\delta } \left[\left(\frac{\delta ^4}{729}+\frac{2 \delta
   }{243}-\frac{7}{729}\right) L_{\bar{\delta }} \right. \nn\\
&&\left.-\frac{5 \delta ^4}{3888}-\frac{19 \delta ^3}{8748}-\frac{7 \delta ^2}{5832}-\frac{25
   \delta }{972}+\frac{\text{Li}_2(\delta )}{243}\right]+\frac{\pi ^2 \delta ^4}{2916}-\frac{5 \delta ^4}{2916}+\frac{247 \delta
   ^3}{34992}-\frac{2461 \delta ^2}{69984}  \nn\\
&& +\frac{\pi ^2 \delta }{486} -\frac{1109 \delta }{11664}-\frac{35 \text{Li}_2(\bar\delta)}{2916}+\frac{7 \text{Li}_2(\delta )}{729}
-\frac{2 \text{Li}_3(\bar \delta )}{243}+\frac{\text{Li}_3(\delta )}{243}+\frac{2 \zeta(3)}{243}+\frac{35 \pi ^2}{17496}
\ ;\nn\\ \\[6mm]
I''(\delta)&=&
-\left(\frac{4 \delta ^4}{729}-\frac{4 \delta ^3}{729}-\frac{4 \delta ^2}{243}+\frac{20 \delta }{729}-\frac{73}{1458}\right)L_{\bar{\delta }}^2
-\left(\frac{L_{\bar{\delta }}}{243}+\frac{\delta ^4}{729}-\frac{\delta ^3}{729}+\frac{\delta^2}{972}+\frac{\delta }{243}\right)L_{\delta }^2 \nn \\
&&+\left(\frac{4 \delta ^4}{243}-\frac{29 \delta ^3}{2187}+\frac{58 \delta ^2}{729}+\frac{5 \delta
   }{2187}+\frac{4 \text{Li}_2(\bar\delta )}{243}+\frac{\pi ^2}{243}-\frac{2215}{17496}\right) L_{\bar{\delta }}
   -L_{\delta} \left[\left(\frac{4 \delta ^4}{729} \right.\right.\nn\\
&&\left.\left.-\frac{4 \delta ^3}{729}+\frac{\delta ^2}{243}+\frac{4 \delta }{243}-\frac{5}{243}\right) L_{\bar{\delta }}
-\frac{2 \delta ^4}{243}+\frac{29 \delta ^3}{4374}-\frac{8 \delta ^2}{729}-\frac{25 \delta }{486}+\frac{2\text{Li}_2(\delta)}{243}\right]
+\frac{\pi^2\delta ^4}{729} \nn\\
&&
-\frac{383 \delta ^4}{23328}-\frac{\pi ^2 \delta ^3}{729}+\frac{923
   \delta ^3}{17496}+\frac{\pi ^2 \delta ^2}{972}-\frac{1357 \delta ^2}{11664}+\frac{\pi ^2 \delta }{243}-\frac{3115 \delta
   }{17496}-\frac{11 \text{Li}_2(\bar\delta )}{486}+\frac{5 \text{Li}_2(\delta )}{243}\nn\\
&&-\frac{4 \text{Li}_3(\bar\delta )}{243}+\frac{2\text{Li}_3(\delta )}{243}+\frac{4 \zeta (3)}{243}+\frac{11 \pi ^2}{2916}\ ;
}
}

While our calculation provides \emph{exactly} all the functions $B^{(\prime\prime)}(z_q,\delta)$, $C^{(\prime,\prime\prime)}(z_q,\delta)$ and
$E^{(\prime\prime)}(z_q,\delta)$, the corresponding expressions depend on $z_q$ and the photon energy $E_\gamma$ through complex functions of
harmonic polylogarithms of various weights, which must be integrated in the region $2 E_\gamma/m_B \in [1-\delta,1]$. Solving these integrals
analytically is highly non-trivial, and even the numerical integration is computationally demanding. We have performed a numerical evaluation of
such integrals and find it more convenient to present the results as an expansion in $\delta$ around the value $\delta=0.316$, and as an interpolation in $z_q$.
These interpolations coincide with the exact results in the region $z\in [0,1]$ to a very good precision. The relevant functions are written as:
\eqa{
{\cal G}(z_q,\delta)
&=& \big[f_{\cal G}^{(0)}(z_q) + i\,h_{\cal G}^{(0)}(z_q)\big] + \big[f_{\cal G}^{(1)}(z_q) + i\,h_{\cal G}^{(1)}(z_q)\big] \,(\delta-0.316)\nn\\[3mm]
&& +\ \big[f_{\cal G}^{(2)}(z_q) + i\,h_{\cal G}^{(2)}(z_q)\big] \,(\delta-0.316)^2+ \cdots
}
with ${\cal G} = B,B'',C,C',C'',E,E''$. The functions $f_{\cal G}^{(i)}(z_q)$, $h_{\cal G}^{(i)}(z_q)$ are fitted to pad\'e approximants of order $[5/5]$:
\eq{
f_{\cal G}^{(i)}(z_q)
= \frac{f_{{\cal G},\{5\}}^{(i)} z_q^5+f_{{\cal G},\{4\}}^{(i)} z_q^4+f_{{\cal G},\{3\}}^{(i)} z_q^3+f_{{\cal G},\{2\}}^{(i)} z_q^2+f_{{\cal G},\{1\}}^{(i)} z_q+f_{{\cal G},\{0\}}^{(i)}}
{f_{{\cal G},\{-5\}}^{(i)} z_q^5+f_{{\cal G},\{-4\}}^{(i)} z_q^4+f_{{\cal G},\{-3\}}^{(i)} z_q^3+f_{{\cal G},\{-2\}}^{(i)} z_q^2+f_{{\cal G},\{-1\}}^{(i)} z_q+1}
}
and similar for $h_{\cal G}^{(i)}(z_q)$. The Pad\'e coefficients $f_{{\cal G},\{j\}}^{(i)},h_{{\cal G},\{j\}}^{(i)}$ are given in
Tables~\ref{tab:coefsB}-\ref{tab:coefsEpp}.


\begin{table}
\small
\ra{1.}
\rb{3.7mm}
\refstepcounter{table}
\label{tab:coefsB}
{\textsf{\small \hspace{4mm} Table \arabic{table}. Pad\'e coefficients for $B(z_q,\delta)$.}}
\vspace{-2mm}
\begin{center}
\begin{tabular}{@{}ccccccc@{}}
\toprule[1.2pt]
$\{j\}$ &  $f_{B,\{j\}}^{(0)}$ &  $f_{B,\{j\}}^{(1)}$ &  $f_{B,\{j\}}^{(2)}$ &  $h_{B,\{j\}}^{(0)}$ &  $h_{B,\{j\}}^{(1)}$ &  $h_{B,\{j\}}^{(2)}$ \\
\midrule[1.1pt]
\{5\}  & 2.6085e2 & -2.3748e3 & 7.8427e2 & -2.5964e0 & -1.8780e1 & -7.0839e1 \\ 
\{4\}  & 5.5417e2 & 1.6248e2 & 8.9587e2 & 3.6935e0 & 2.6299e1 & 8.9418e1 \\ 
\{3\}  & -8.6141e0 & 1.2216e3 & -2.0894e2 & -1.6902e0 & -1.1864e1 & -3.5637e1 \\ 
\{2\}  & -1.5107e0 & -9.2378e1 & 2.1731e1 & 3.2301e-1 & 2.2545e0 & 6.0307e0 \\ 
\{1\}  & 1.1241e-1 & 5.4522e0 & -1.3841e0 & -2.6875e-2 & -1.8840e-1 & -4.5543e-1 \\ 
\{0\}  & 3.2101e-3 & 2.0675e-2 & 4.0727e-2 & 8.0387e-4 & 5.7122e-3 & 1.2664e-2 \\ 
\{-1\}  & 2.7478e1 & 2.4959e2 & -3.5865e1 & 5.1543e0 & -9.6139e0 & -2.1925e1 \\ 
\{-2\}  & -6.5543e1 & -4.3377e3 & 5.9242e2 & -2.5093e2 & -1.8937e2 & -1.6101e2 \\ 
\{-3\}  & -9.6131e3 & 5.7863e4 & -5.6283e3 & 1.7465e4 & 1.1349e4 & 1.1056e4 \\ 
\{-4\}  & 2.2612e5 & 7.7096e4 & 2.1131e4 & -4.3025e5 & -2.3262e5 & -1.5200e5 \\ 
\{-5\}  & 1.6782e5 & -2.0939e5 & 3.8351e4 & 4.3797e6 & 1.7123e6 & 7.2442e5 \\ 
\bottomrule[1.2pt]
\end{tabular}
\vspace{0mm}
\end{center}
\end{table}


\begin{table}
\small
\ra{1}
\rb{3.5mm}
\refstepcounter{table}
\label{tab:coefsBpp}
{\textsf{\small \hspace{4mm} Table \arabic{table}. Pad\'e coefficients for $B''(z_q,\delta)$.}}
\vspace{-2mm}
\begin{center}
\begin{tabular}{@{}ccccccc@{}}
\toprule[1.2pt]
$\{j\}$ &  $f_{B'',\{j\}}^{(0)}$ &  $f_{B'',\{j\}}^{(1)}$ &  $f_{B'',\{j\}}^{(2)}$ &  $h_{B'',\{j\}}^{(0)}$ &  $h_{B'',\{j\}}^{(1)}$ &  $h_{B'',\{j\}}^{(2)}$ \\
\midrule[1.1pt]
\{5\}  & -9.5739e-1 & -9.6422e0 & -2.8386e1 & -7.9067e-2 & -6.6893e-1 & -2.9051e0 \\ 
\{4\}  & 3.0239e0 & 2.7036e1 & 1.2991e2 & 1.9267e-1 & 1.6095e0 & 6.3544e0 \\ 
\{3\}  & -3.9058e-1 & -6.8880e-1 & 2.3899e0 & -1.7623e-1 & -1.4512e0 & -5.1486e0 \\ 
\{2\}  & 3.7847e-3 & -6.4972e-1 & -4.0154e0 & 7.5813e-2 & 6.1479e-1 & 1.9516e0 \\ 
\{1\}  & -9.5759e-3 & 3.6265e-3 & 2.4684e-1 & -1.5417e-2 & -1.2304e-1 & -3.5068e-1 \\ 
\{0\}  & 3.9641e-3 & 3.0018e-2 & 7.5378e-2 & 1.1925e-3 & 9.3670e-3 & 2.4204e-2 \\ 
\{-1\}  & -4.6336e0 & -2.3791e0 & 2.8343e-1 & -7.6041e0 & -6.7239e0 & -5.9052e0 \\ 
\{-2\}  & 3.1991e1 & 1.5265e1 & -1.9554e0 & 5.0750e1 & 4.1512e1 & 2.0995e1 \\ 
\{-3\}  & -2.6083e2 & -2.3013e2 & -3.0385e2 & -1.6999e2 & -8.9622e1 & 2.8933e2 \\ 
\{-4\}  & 1.1184e3 & 1.4088e3 & 2.6697e3 & -2.6338e2 & -7.5669e2 & -4.3799e3 \\ 
\{-5\}  & -3.1470e2 & -4.4778e2 & -4.4635e2 & 3.2029e3 & 5.1645e3 & 1.9822e4 \\ 
\bottomrule[1.2pt]
\end{tabular}
\vspace{4mm}
\end{center}
%
%
\ra{1}
\rb{3.9mm}
\refstepcounter{table}
\label{tab:coefsC}
{\textsf{\small \hspace{4mm} Table \arabic{table}. Pad\'e coefficients for $C(z_q,\delta)$.}}
\vspace{-2mm}
\begin{center}
\begin{tabular}{@{}ccccccc@{}}
\toprule[1.2pt]
$\{j\}$ &  $f_{C,\{j\}}^{(0)}$ &  $f_{C,\{j\}}^{(1)}$ &  $f_{C,\{j\}}^{(2)}$ &  $h_{C,\{j\}}^{(0)}$ &  $h_{C,\{j\}}^{(1)}$ &  $h_{C,\{j\}}^{(2)}$ \\
\midrule[1.1pt]
\{5\}  & 6.7269e2 & 4.9426e2 & 1.0451e3 & -3.4512e1 & -2.3670e2 & -9.4447e2 \\ 
\{4\}  & 3.3679e3 & 5.8344e3 & 3.4974e3 & 4.8342e1 & 3.2457e2 & 1.1315e3 \\ 
\{3\}  & 2.3178e2 & 2.4369e2 & -6.8258e2 & -2.1749e1 & -1.4292e2 & -4.2386e2 \\ 
\{2\}  & -3.3437e1 & -8.0765e1 & 7.0897e1 & 4.0984e0 & 2.6571e1 & 6.7645e1 \\ 
\{1\}  & 1.7801e0 & 2.8034e0 & -5.3246e0 & -3.3765e-1 & -2.1805e0 & -4.8418e0 \\ 
\{0\}  & 1.8570e-2 & 1.0740e-1 & 1.8133e-1 & 1.0040e-2 & 6.5154e-2 & 1.2819e-1 \\ 
\{-1\}  & 7.1021e1 & 1.0292e1 & -3.4108e1 & 9.6116e0 & -8.1610e0 & -2.2268e1 \\ 
\{-2\}  & -6.9880e2 & -1.5360e2 & 5.3644e2 & -6.8385e2 & -3.9083e2 & -2.5150e2 \\ 
\{-3\}  & -5.9528e3 & -6.3499e3 & -5.0770e3 & 3.0179e4 & 1.5933e4 & 1.3515e4 \\ 
\{-4\}  & 3.3626e5 & 9.9501e4 & 2.0032e4 & -6.1404e5 & -2.6974e5 & -1.6966e5 \\ 
\{-5\}  & 1.9465e5 & 4.3193e4 & 3.0917e4 & 5.0960e6 & 1.7175e6 & 7.4113e5 \\ 
\bottomrule[1.2pt]
\end{tabular}
\vspace{4mm}
\end{center}
%
%
\ra{1}
\rb{3.7mm}
\refstepcounter{table}
\label{tab:coefsCp}
{\textsf{\small \hspace{4mm} Table \arabic{table}. Pad\'e coefficients for $C'(z_q,\delta)$.}}
\vspace{-2mm}
\begin{center}
\begin{tabular}{@{}ccccccc@{}}
\toprule[1.2pt]
$\{j\}$ &  $f_{C',\{j\}}^{(0)}$ &  $f_{C',\{j\}}^{(1)}$ &  $f_{C',\{j\}}^{(2)}$ &  $h_{C',\{j\}}^{(0)}$ &  $h_{C',\{j\}}^{(1)}$ &  $h_{C',\{j\}}^{(2)}$ \\
\midrule[1.1pt]
\{5\}  & 3.7841e2 & 5.9118e2 & 3.9056e2 & -4.5121e-1 & -6.0463e1 & -2.1698e1 \\ 
\{4\}  & 2.7044e2 & 4.8884e2 & 5.4690e2 & 7.9268e-1 & 8.8873e1 & 3.5043e1 \\ 
\{3\}  & -1.1579e2 & -1.8312e2 & -1.8791e2 & -5.0952e-1 & -3.9767e1 & -1.8854e1 \\ 
\{2\}  & 1.3594e1 & 1.9867e1 & 2.5682e1 & 1.5463e-1 & 5.9243e0 & 4.2454e0 \\ 
\{1\}  & 3.0297e-2 & -1.6165e-1 & -1.7434e0 & -2.2768e-2 & -1.7640e-1 & -4.0900e-1 \\ 
\{0\}  & 5.8184e-3 & 3.5538e-2 & 5.7703e-2 & 1.3344e-3 & 8.5805e-3 & 1.5244e-2 \\ 
\{-1\}  & 1.4953e0 & -7.2811e0 & -3.1995e1 & 1.1886e1 & -8.8975e0 & -2.7485e1 \\ 
\{-2\}  & 2.4503e3 & 6.0955e2 & 4.7985e2 & -5.2688e2 & 4.2689e2 & 3.3788e2 \\ 
\{-3\}  & -1.9815e4 & -5.2682e3 & -3.4156e3 & 5.3900e3 & 1.1219e4 & -9.5618e2 \\ 
\{-4\}  & 3.7632e4 & 1.1781e4 & 8.8305e3 & -2.3444e4 & -1.4783e5 & -1.2086e4 \\ 
\{-5\}  & 1.0249e5 & 2.6408e4 & 1.1136e4 & 3.8372e4 & 5.4497e5 & 8.0691e4 \\ 
\bottomrule[1.2pt]
\end{tabular}
\end{center}
\end{table}
%


\begin{table}
\small
\ra{1.}
\rb{3.8mm}
\refstepcounter{table}
\label{tab:coefsCpp}
{\textsf{\small \hspace{4mm} Table \arabic{table}. Pad\'e coefficients for $C''(z_q,\delta)$.}}
\vspace{-2mm}
\begin{center}
\begin{tabular}{@{}ccccccc@{}}
\toprule[1.2pt]
$\{j\}$ &  $f_{C'',\{j\}}^{(0)}$ &  $f_{C'',\{j\}}^{(1)}$ &  $f_{C'',\{j\}}^{(2)}$ &  $h_{C'',\{j\}}^{(0)}$ &  $h_{C'',\{j\}}^{(1)}$ &  $h_{C'',\{j\}}^{(2)}$ \\
\midrule[1.1pt]
\{5\}  & -7.7072e0 & -7.3684e1 & -3.2672e2 & -8.7364e-1 & -6.3940e0 & -3.5367e1 \\ 
\{4\}  & 6.5728e0 & 3.6241e1 & 1.4444e2 & 2.0978e0 & 1.5469e1 & 7.3871e1 \\ 
\{3\}  & 4.6807e0 & 7.0186e1 & 3.8320e2 & -1.8917e0 & -1.4054e1 & -5.6847e1 \\ 
\{2\}  & -1.5486e0 & -1.9670e1 & -9.3766e1 & 8.0372e-1 & 6.0118e0 & 2.0476e1 \\ 
\{1\}  & 1.0573e-1 & 1.6449e0 & 8.0166e0 & -1.6179e-1 & -1.2175e0 & -3.5220e0 \\ 
\{0\}  & 1.8873e-2 & 1.3658e-1 & 3.3186e-1 & 1.2423e-2 & 9.3960e-2 & 2.3583e-1 \\ 
\{-1\}  & -4.4964e-1 & 4.8147e0 & 1.4577e1 & -6.6212e0 & -4.6773e0 & -2.9278e0 \\ 
\{-2\}  & -2.5309e0 & -4.5104e1 & -1.3130e2 & 2.2861e1 & -8.8411e0 & -7.8932e1 \\ 
\{-3\}  & -1.7756e2 & -5.9658e1 & 1.1661e2 & 1.0154e2 & 4.6805e2 & 1.6895e3 \\ 
\{-4\}  & 1.3958e3 & 1.8710e3 & 3.9334e3 & -1.5110e3 & -3.7935e3 & -1.3934e4 \\ 
\{-5\}  & -1.0185e3 & -1.5684e3 & -3.2201e3 & 5.1086e3 & 1.1386e4 & 4.4771e4 \\ 
\bottomrule[1.2pt]
\end{tabular}
\vspace{4mm}
\end{center}
%
%
\ra{1.}
\rb{3.4mm}
\refstepcounter{table}
\label{tab:coefsE}
{\textsf{\small \hspace{4mm} Table \arabic{table}. Pad\'e coefficients for $E(z_q,\delta)$.}}
\vspace{-2mm}
\begin{center}
\begin{tabular}{@{}ccccccc@{}}
\toprule[1.2pt]
$\{j\}$ &  $f_{E,\{j\}}^{(0)}$ &  $f_{E,\{j\}}^{(1)}$ &  $f_{E,\{j\}}^{(2)}$ &  $h_{E,\{j\}}^{(0)}$ &  $h_{E,\{j\}}^{(1)}$ &  $h_{E,\{j\}}^{(2)}$ \\
\midrule[1.1pt]
\{5\}  & 6.4806e1 & 3.5981e2 & 1.2940e2 & 2.1801e1 & 1.5065e2 & 6.3214e2 \\ 
\{4\}  & -8.5279e2 & -1.5483e3 & -1.4843e3 & -3.0643e1 & -2.0732e2 & -7.5761e2 \\ 
\{3\}  & -9.9349e1 & -4.0637e2 & 2.2130e2 & 1.3820e1 & 9.1544e1 & 2.8342e2 \\ 
\{2\}  & 1.2871e1 & 6.1243e1 & -2.0763e1 & -2.6077e0 & -1.7054e1 & -4.5145e1 \\ 
\{1\}  & -5.9578e-1 & -2.2780e0 & 1.9360e0 & 2.1490e-1 & 1.4015e0 & 3.2245e0 \\ 
\{0\}  & -7.3348e-3 & -4.1945e-2 & -7.8462e-2 & -6.3860e-3 & -4.1912e-2 & -8.5194e-2 \\ 
\{-1\}  & 4.7343e1 & 2.3190e1 & -3.2893e1 & 1.4649e1 & -3.5937e0 & -1.8952e1 \\ 
\{-2\}  & -1.4068e2 & -3.0228e2 & 5.2625e2 & -7.6567e2 & -5.4265e2 & -3.8314e2 \\ 
\{-3\}  & -1.3594e4 & -7.1646e3 & -5.3673e3 & 3.2909e4 & 1.9737e4 & 1.5799e4 \\ 
\{-4\}  & 3.2738e5 & 1.3102e5 & 2.2555e4 & -6.6974e5 & -3.2807e5 & -1.9062e5 \\ 
\{-5\}  & 2.8262e5 & 4.3433e4 & 3.7019e4 & 5.6997e6 & 2.0888e6 & 8.2259e5 \\
\bottomrule[1.2pt]
\end{tabular}
\vspace{4mm}
\end{center}
%
%
\ra{1.}
\rb{3.4mm}
\refstepcounter{table}
\label{tab:coefsEpp}
{\textsf{\small \hspace{4mm} Table \arabic{table}. Pad\'e coefficients for $E''(z_q,\delta)$.}}
\vspace{-2mm}
\begin{center}
\begin{tabular}{@{}ccccccc@{}}
\toprule[1.2pt]
$\{j\}$ &  $f_{E'',\{j\}}^{(0)}$ &  $f_{E'',\{j\}}^{(1)}$ &  $f_{E'',\{j\}}^{(2)}$ &  $h_{E'',\{j\}}^{(0)}$ &  $h_{E'',\{j\}}^{(1)}$ &  $h_{E'',\{j\}}^{(2)}$ \\
\midrule[1.1pt]
\{5\}  & -2.7046e1 & -1.1652e2 & -6.3378e3 & 4.8640e-1 & 3.9114e0 & 1.7243e1 \\ 
\{4\}  & -4.2392e2 & -4.6203e3 & -5.8000e4 & -1.1720e0 & -9.3898e0 & -3.7292e1 \\ 
\{3\}  & 1.3189e2 & 1.3432e3 & 1.5166e4 & 1.0609e0 & 8.4650e0 & 3.0070e1 \\ 
\{2\}  & -8.8744e0 & -9.3948e1 & -1.1660e3 & -4.5251e-1 & -3.5954e0 & -1.1478e1 \\ 
\{1\}  & -1.4275e0 & -1.2464e1 & -8.5821e1 & 9.1466e-2 & 7.2362e-1 & 2.1083e0 \\ 
\{0\}  & -4.5677e-3 & -3.6869e-2 & -1.1199e-1 & -7.0517e-3 & -5.5554e-2 & -1.5106e-1 \\ 
\{-1\}  & 2.8279e2 & 3.0406e2 & 7.1445e2 & -7.6394e0 & -6.0846e0 & -4.0315e0 \\ 
\{-2\}  & -1.2889e3 & -9.3334e2 & 1.1757e3 & 3.3059e1 & 7.5335e0 & -4.7914e1 \\ 
\{-3\}  & 9.6814e3 & 7.3874e3 & -7.4918e3 & 1.2448e1 & 2.7671e2 & 1.2068e3 \\ 
\{-4\}  & -7.8680e4 & -8.4193e4 & -1.9839e5 & -1.0211e3 & -2.5305e3 & -1.0096e4 \\ 
\{-5\}  & 2.9932e5 & 3.9930e5 & 1.7241e6 & 3.8601e3 & 7.8008e3 & 3.2678e4 \\ 
\bottomrule[1.2pt]
\end{tabular}
\vspace{-3mm}
\end{center}
\end{table}


The functions $\widetilde{\cal F}(z_q,\delta)$ are UV finite and collinear safe. Again, we have the following relationship,
\eq{
\widetilde{\cal F}_{(i)}^{s,1}(z_q,\delta) = -6\,\widetilde{\cal F}_{(i)}^{\times,1}(z_q,\delta)\ .
}
As before, the `crossed' functions $\widetilde{\cal F}_{(J)}^{\times,1}(z_q,\delta)$ are known exactly but we provide here simplified expressions
as an expansion in $(\delta-0.316)$ and interpolated in $z_q$. We write:
\eqa{
\widetilde{\cal F}_{(J)}^{\times,1}(z_q,\delta)
&=& \big[\tilde f_{(J)}^{(0)}(z_q) + i\,\tilde h_{(J)}^{(0)}(z_q)\big]
+ \big[\tilde f_{(J)}^{(1)}(z_q) + i\,\tilde h_{(J)}^{(1)}(z_q)\big] \,(\delta-0.316)\nn\\[3mm]
&& +\ \big[\tilde f_{(J)}^{(2)}(z_q) + i\,\tilde h_{(J)}^{(2)}(z_q)\big] \,(\delta-0.316)^2+ \cdots
}
with $J=i,ii$. The functions $\tilde f_{(J)}^{(i)}(z_q)$ are again fitted to Pad\'e approximants:
{\small
\eqa{
\tilde f_{(J)}^{(i)}(z_q)
&=& \frac{\tilde f_{(J),\{5\}}^{(i)} z_q^5+\tilde f_{(J),\{4\}}^{(i)} z_q^4+\tilde f_{(J),\{3\}}^{(i)} z_q^3+\tilde f_{(J),\{2\}}^{(i)} z_q^2+\tilde f_{(J),\{1\}}^{(i)} z_q+\tilde f_{(J),\{0\}}^{(i)}}
{\tilde f_{(J),\{-6\}}^{(i)} z_q^6+\tilde f_{(J),\{-5\}}^{(i)} z_q^5+\tilde f_{(J),\{-4\}}^{(i)} z_q^4+\tilde f_{{(J)},\{-3\}}^{(i)} z_q^3+\tilde f_{(J),\{-2\}}^{(i)} z_q^2+\tilde f_{(J),\{-1\}}^{(i)} z_q+1}\nn\\[2mm]
}
}
but a different parameterization for the functions $\tilde h_{(J)}^{(i)}(z_q)$ is found to reproduce the exact result more accurately. While for $\tilde h_{(J)}^{(0)}(z_q)$ and $\tilde h_{(J)}^{(1)}(z_q)$ we use
\begin{align}
\tilde h_{(J)}^{(i)}(z_q)&= z_q \, \exp\big[-\tilde h_{(J),\{e\}}^{(i)} \, z_q\big] \, (\frac{1}{4}-z_q)^2 \, \theta(\frac{1}{4}-z_q) \, \sum_{j=0}^{6} \tilde h_{(J),\{j\}}^{(i)} \, z_q^j \, , \\
\intertext{we make the ansatz}
\tilde h_{(J)}^{(2)}(z_q)&= z_q \, (\frac{1}{4}-z_q)^2 \, \theta(\frac{1}{4}-z_q) \, \frac{\sum_{j=0}^{7} \tilde h_{(J),\{j\}}^{(2)} \, z_q^j}{1+\sum_{j=1}^{7} \tilde h_{(J),\{j\}}^{(2)} \, z_q^j}
\end{align}
for $\tilde h_{(J)}^{(2)}(z_q)$. The coefficients $\tilde f_{(J),\{j\}}^{(i)}$ and $\tilde h_{(J),\{j\}}^{(i)}$ can be found in Tables~\ref{tab:coefsFtildereal} and~\ref{tab:coefsFtildeimag}, respectively.
%
%
\begin{table}
\small
\ra{1.}
\rb{3.6mm}
\refstepcounter{table}
\label{tab:coefsFtildereal}
{\textsf{\small \hspace{4mm} Table \arabic{table}. Pad\'e coefficients for the real parts of $\widetilde {\cal F}_{(J)}^{\times,1} (z_q,\delta)$.}}
\vspace{-2mm}
\begin{center}
\begin{tabular}{@{}ccccccc@{}}
\toprule[1.2pt]
$\{j\}$ &  $\tilde f_{(i),\{j\}}^{(0)}$ &  $\tilde f_{(i),\{j\}}^{(1)}$ &  $\tilde f_{(i),\{j\}}^{(2)}$ &  $\tilde f_{(ii),\{j\}}^{(0)}$ &  $\tilde f_{(ii),\{j\}}^{(1)}$ &  $\tilde f_{(ii),\{j\}}^{(2)}$ \\
\midrule[1.1pt]
\{5\}  & -6.0396e1 & -7.9313e2 & -4.1781e2 & 8.5226e3 & 5.9559e4 & 6.0577e2 \\ 
\{4\}  & -4.3337e1 & 4.3943e1 & 1.7075e2 & 5.2397e2 & -1.7693e3 & -2.4607e2 \\ 
\{3\}  & 2.7398e1 & 7.1810e1 & -2.3013e1 & -6.9260e2 & -2.5828e3 & 4.0981e1 \\ 
\{2\}  & -4.9232e0 & -1.6196e1 & 1.0162e0 & 1.3412e2 & 5.3783e2 & -3.9970e0 \\ 
\{1\}  & 1.9455e-1 & 6.9739e-1 & -1.0853e-2 & -3.4138e0 & -1.4094e1 & 2.4014e-1 \\ 
\{0\}  & 4.7105e-4 & 2.1302e-3 & 4.5410e-4 & -1.5811e-3 & -8.3330e-3 & -5.1000e-3 \\ 
\{-1\}  & 3.2507e2 & 2.5362e2 & -6.7339e1 & 1.8946e3 & 1.4787e3 & -3.6244e1 \\ 
\{-2\}  & -1.2129e3 & -3.3619e3 & 2.9914e3 & 2.7983e4 & 1.4145e4 & 1.3530e3 \\ 
\{-3\}  & 4.5871e4 & 3.7550e4 & -6.7242e4 & 3.3450e5 & -3.5081e4 &-3.2312e4 \\ 
\{-4\}  & -2.7172e5 & 8.5735e4 & 7.6883e5 & 2.2367e6 & 7.2362e6 & 4.1425e5 \\ 
\{-5\}  & -3.6784e5 & -2.6336e6 & -4.2134e6 & -4.8970e7 & -7.8826e7 & -2.5466e6 \\
\{-6\}  & 4.8301e6 & 9.5241e6 & 8.8361e6 & 3.1279e8 & 4.0127e8 & 5.9475e6 \\
\bottomrule[1.2pt]
\end{tabular}
\vspace{4mm}
\end{center}
%
%
%
\ra{1.}
\rb{3.5mm}
\refstepcounter{table}
\label{tab:coefsFtildeimag}
{\textsf{\small \hspace{2mm} Table \arabic{table}. Coefficients for the imaginary parts of $\widetilde {\cal F}_{(J)}^{\times,1} (z_q,\delta)$.}}
\vspace{-31mm}

\begin{minipage}{.65\textwidth}
\begin{center}
\begin{tabular}{@{}ccccc@{}}
\toprule[1.2pt]
$\{j\}$ &  $\tilde h_{(i),\{j\}}^{(0)}$ &  $\tilde h_{(i),\{j\}}^{(1)}$ & $\tilde h_{(ii),\{j\}}^{(0)}$ &  $\tilde h_{(ii),\{j\}}^{(1)}$ \\
\midrule[1.1pt]
\{6\} & 0  & 0 & 2.7269e9 & 3.0740e10 \\ 
\{5\} & 0  &  5.9601e5 & -5.6512e8 & -5.4396e9 \\ 
\{4\} & -3.4279e4  & -2.5773e5  & 5.1876e7 & 4.4076e8 \\ 
\{3\} & 5.4423e3  & 3.5916e4  & -2.2643e6 & -1.6493e7 \\ 
\{2\} & -5.0377e2  & -2.5068e3  & 5.7546e4 & 3.7561e5 \\ 
\{1\} & 9.7190e0  &  2.9523e1 & -5.7810e2 & -3.0581e3 \\ 
\{0\} & -7.3742e-1  & -2.5466e0  & 6.7661e0 & 3.4673e1 \\ 
\{$e$\} & 2.9801e1   & 2.3898e1 & 8.2260e1 & 9.3516e1 \\
\bottomrule[1.2pt]
\end{tabular}
\vspace{4mm}
\end{center}
\end{minipage}
%
%
\vspace{-2mm}
\begin{minipage}{.35\textwidth}
\vspace{32mm}
\begin{center}
\begin{tabular}{@{}ccc@{}}
\toprule[1.2pt]
$\{j\}$ &  $\tilde h_{(i),\{j\}}^{(2)}$ &  $\tilde h_{(ii),\{j\}}^{(2)}$ \\
\midrule[1.1pt]
\{7\}  & 1.9156e7 & 0 \\
\{6\}  & -1.2045e7 & -4.0614e7 \\ 
\{5\}  & 2.7691e6 & 2.8773e7 \\ 
\{4\}  & -2.4695e5 & -9.2094e6 \\ 
\{3\}  & -1.3034e3 & 1.6547e6 \\ 
\{2\}  & 1.5514e3 & -1.5575e5 \\ 
\{1\}  & -8.3143e1 & 5.4749e3 \\ 
\{0\}  & 1.1641e0 & 3.0407e1 \\ 
\{-1\}  & 5.3853e1 & 1.0191e3\\
\{-2\}  & -4.7202e3 & 5.4442e3\\ 
\{-3\}  & 1.2537e5 & -6.7695e5\\ 
\{-4\}  & -1.7062e6 & 1.0194e7\\ 
\{-5\}  & 1.2920e7 & -4.8380e7\\ 
\{-6\}  & -5.1347e7 & -2.3921e7\\ 
\{-7\}  & 8.3239e7 & 4.9141e8\\ 
\bottomrule[1.2pt]
\end{tabular}
\vspace{4mm}
\end{center}
\end{minipage}
\end{table}
Finally the functions $\widetilde{\cal F}^{I,4}_{(J)}(\delta)$ are given by
\eq{
\widetilde{\cal F}_{(i)}^{s,4}(\delta) = -6\,\widetilde{\cal F}_{(i)}^{\times,4}(\delta)
}
and
\eqa{
\widetilde{\cal F}_{(i)}^{\times,4}(\delta) =  -0.0000513772
-0.0003375398 \,(\delta-0.316) - 0.000532746\,(\delta-0.316)^2 + \cdots \, , \nn \\[2mm]
\widetilde{\cal F}_{(ii)}^{\times,4}(\delta) = -0.0001176336 - 0.0003362453 \,(\delta-0.316)
+ 0.001067501\,(\delta-0.316)^2+ \cdots \, . \nn \\
}



\end{document}